\documentclass[twocolumn,twoside]{IEEEtran}

\ifCLASSINFOpdf

\else

\fi

\ifCLASSOPTIONcompsoc
    \usepackage[caption=false, font=normalsize, labelfont=sf, textfont=sf]{subfig}
\else
    \usepackage[caption=false, font=normalsize]{subfig}
\fi
\usepackage{lipsum}%
\usepackage[dvipsnames]{xcolor}

\usepackage{balance}
\usepackage{multicol}   
\usepackage{cite}
\usepackage{gensymb}
\usepackage{multirow}
\usepackage{graphics}
\usepackage{epsfig}
\usepackage{graphicx}
\usepackage{epstopdf}
\usepackage{textcomp}
\usepackage{amsmath}
\usepackage{mathtools}
\usepackage{filecontents}
\usepackage{lipsum,color}
\usepackage{amssymb}
\usepackage{float}
\usepackage{times} 
\usepackage{amsthm}  

\usepackage{amsfonts}

\theoremstyle{break}

\begin{document}
\title{Analytical Channel Models for Millimeter Wave UAV Networks under Hovering Fluctuations}


\author{Mohammad~Taghi~Dabiri,~Hossein~Safi,~Saeedeh~Parsaeefard,~{\it Senior Member,~IEEE},\\~and~Walid~Saad,~{\it Fellow, IEEE}}

\maketitle

\begin{abstract}
%
The integration of unmanned aerial vehicles (UAVs) and millimeter wave (mmWave) wireless systems
has been recently proposed to provide high data rate
aerial links for next generation wireless networks.
However, establishing UAV-based mmWave links is quite
challenging due to the random fluctuations of hovering UAVs which can induce antenna gain mismatch between transmitter and receiver.
To assess the benefit of UAV-based mmWave links, in this paper, tractable, closed-form statistical channel models are derived for three UAV communication scenarios: (i) a direct UAV-to-UAV link, (ii) an aerial relay link in which source, relay, and destination are hovering UAVs, and (iii) a relay link in which a hovering UAV connects a ground source to a ground destination.
The accuracy of the derived analytical expressions is corroborated by performing Monte-Carlo simulations. Numerical results are then used to study the effect of antenna directivity gain under different channel conditions for establishing reliable UAV-based  mmWave links in terms of achieving minimum outage probability.
It is shown that the performance of such links is largely
dependent on the random fluctuations of hovering UAVs. Moreover, higher antenna
directivity gains achieve better performance at low SNR
regime. Nevertheless, at the high SNR regime, lower antenna directivity gains result in a more reliable communication link. The developed results can therefore be applied as a benchmark for finding the optimal antenna directivity gain of UAVs under the different levels of instability without resorting to time-consuming simulations.

\end{abstract}
\IEEEpeerreviewmaketitle

\section{Introduction}
\begin{figure}
	\begin{center}
		\includegraphics[width=3.35 in]{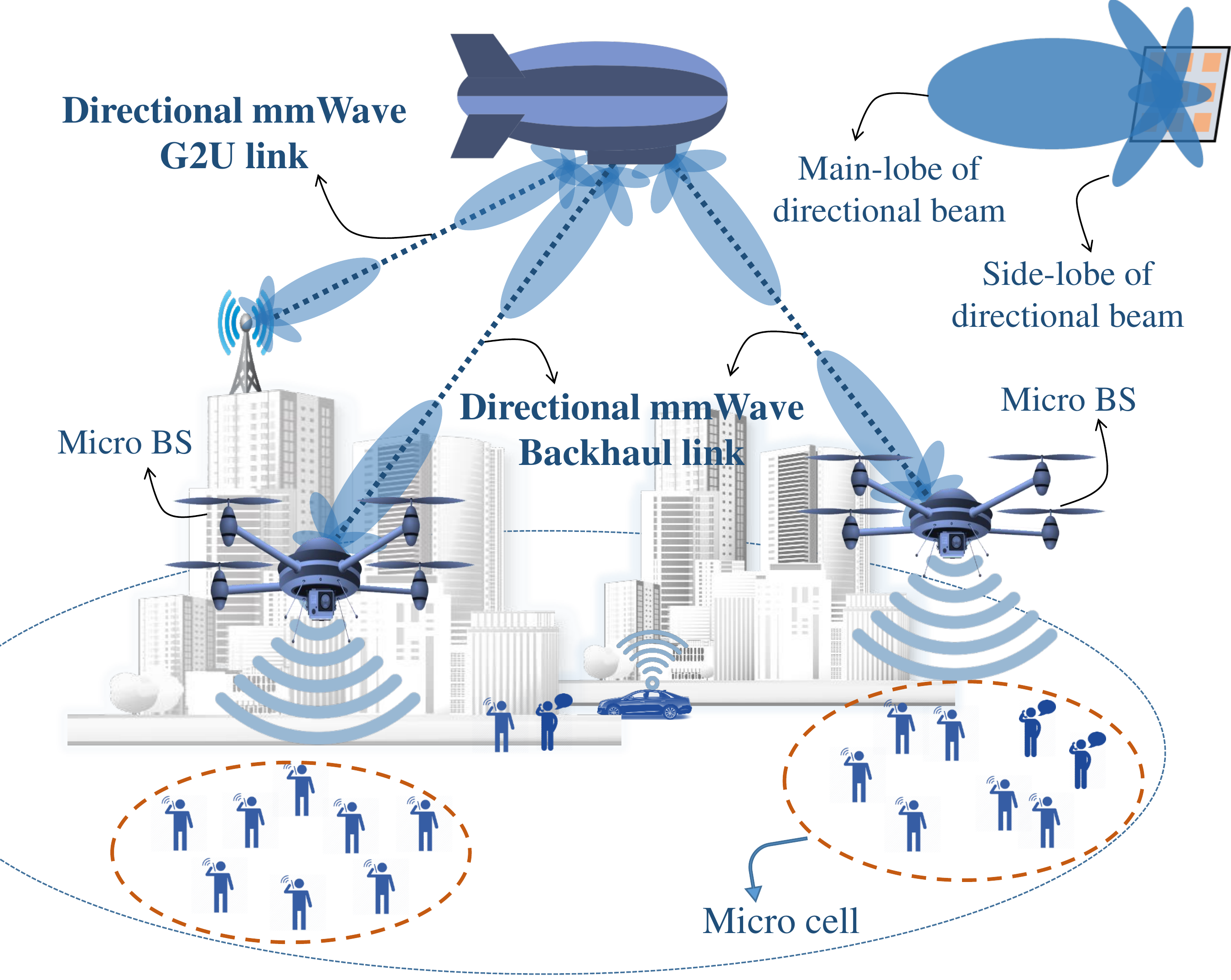}
		\caption{Nominal illustration of the next generation of mobile networks employing air-to-air and air-to-ground links to provide ubiquitous coverage. Directional mmWave links are highlighted as backhaul of such UAV-assisted networks. }
		\label{aerial-BSs}
	\end{center}
\end{figure}

Unmanned aerial vehicles (UAVs) are seen as an important feature of next-generation wireless cellular systems \cite{saad2019vision,mozaffari2018tutorial,mozaffari2016unmanned}.
Due to their unique capabilities such as maneuverability, flexibility, and adaptive altitude adjustment, UAVs are able to be used in different situations, especially they can use as aerial base-stations (BSs) to provide the ubiquitous connectivity for the next generation of wireless networks\cite{bor2016new,zhang2019research}.
As seen in Fig. \ref{aerial-BSs}, cellular networks can be equipped with UAVs as flying BSs or mobile relay backhaul nodes which effectively enhance the coverage of heterogeneous networks. UAVs can also be seen as prime candidates to support mmWave communications.
Owing to the small wavelength at mmWave frequency, the integration of beam-steerable directive antenna arrays on a small UAV with limited payload is a promising solution to provide high-capacity communication links as discussed in \cite{zhao2018channel} and \cite{yu2018capacity}. Indeed, in order to fully exploit the benefit of mmWave frequencies, the receiver must be placed within the line-of-sight (LoS) of the transmitter \cite{wang2018mmwave}. Due to the impracticality of establishing long LoS links, especially in dense urban environments, UAVs equipped with mmWave capabilities are able to provide on-the-fly communications and establish reliable LoS links between each other and also the ground station \cite{khan2018line}.


%

However,  to reap the benefits of mmWave-enabled UAV communications, the communication channel for airborne mmWave links should be distinctively characterized in terms of a UAV's random vibrations due to hovering fluctuations and also mmWave propagation characteristics. Moreover, due to the directionality of the mmWave transmit pattern,  UAV-based mmWave communication systems suffer from misalignment between transceivers \cite{choi2016millimeter}. In particular, random vibrations of UAVs can lead to antenna gain mismatch between transmitter and receiver which, in turn, can cause signal-to-noise ratio (SNR) fluctuations at the receiver side that can significantly degrade the reliability of the system.
To avoid possible gain mismatch between transceivers and have a reliable mmWave UAV link under different levels of UAV fluctuations, the antenna radiation pattern of transceivers should be optimally designed.
In fact, optimizing the radiation pattern requires balancing an inherent tradeoff between increasing directional gain to compensate for the large path loss at mmWave frequencies and decreasing it to alleviate the adverse effect of transceivers vibrations.
Therefore, to quantify the benefits of using mmWave frequencies for UAV-based communication systems, it is essential to have an accurate channel modeling that incorporates the effect of path loss and fading of mmWave signals along with the effect of UAV fluctuations on directional antenna.

\subsection{Literature Review}
{There has been a surge of recent works on the use of UAVs for communications \cite{chen2017caching,mozaffari2019beyond,mozaffari2019communications,khoshkholgh2019non,bor20195g}.
To assess the benefits of UAV deployment at mmWave frequencies, it is important to have a comprehensive and accurate channel model while taking into account mmWave signal propagation  as well as the random effect of hovering fluctuations.
For instance, even though channel modeling in the context of UAV communications has been discussed in recent studies \cite{al2014optimal,matolak2017air,sun2017air}, these prior works are restricted to sub-6 GHz bands whose results cannot directly extended to mmWave systems.
Meanwhile, most of the prior art on mmWave communications \cite{semiari2018caching,akdeniz2014millimeter,wu201760,sulyman2014radio} does not address the presence of UAVs, with the exception of a few recent works in \cite{khawaja2017uav,khawaja2018temporal,xiao2016enabling,kovalchukov2018analyzing,rupasinghe2019non,galkin2018backhaul,gapeyenko2018flexible}.}

For instance, in \cite{khawaja2017uav}, the authors propose a ray tracing approach to characterize the mmWave propagation channel for an air-to-ground link  at  28 GHz and 60 GHz.
Meanwhile, the work in \cite{khawaja2018temporal} studies the small scale temporal and spatial characteristics of mmWave air-to-ground LoS propagation channels at 28 GHz under different conditions.
However, the authors in \cite{khawaja2018temporal} assumed half-wave dipole antennas with an omni-directional pattern and, hence, their approach does not properly capture the UAV fluctuation effects.
The work in \cite{xiao2016enabling} analyzes the characteristics of directional mmWave wireless channel propagation characteristics for UAV cellular networks  by considering the Doppler effect as a result of UAV movement.
In \cite{kovalchukov2018analyzing}, the authors propose a novel three-dimensional model for UAV-based mmWave communication that captures the high directionality of transmissions as well as the UAV random heights.
In \cite{rupasinghe2019non}, the authors introduce an analytical framework for non-orthogonal multiple-access transmission at UAV-based BS to serve more users simultaneously in a hot spot area such as a football stadium.
However, the proposed channel models in \cite{xiao2016enabling,kovalchukov2018analyzing,rupasinghe2019non} are only applicable for a link between users and a perfectly stable UAV.
In \cite{galkin2018backhaul}, the authors use stochastic geometry to study directional UAV-based mmWave backhaul links operating at 73 GHz and then, compare their performance with an LTE backhaul operating at 2 GHz. For simplicity, in \cite{galkin2018backhaul}, the antenna pattern is approximated by a rectangular radiation pattern.
In \cite{gapeyenko2018flexible}, the authors propose a flexible UAV-assisted backhaul link that takes into account the dynamic blockage of mmWave links.
However, the works in \cite{galkin2018backhaul,gapeyenko2018flexible} ignore the random fluctuations of UAVs and assume a perfect beam alignment between transceivers.
Obviously, because of  the random fluctuations as well as strict payload limitations for employing antenna stabilizers, careful alignment is not practically feasible in aerial links (particularly, for small multi-rotor UAVs) which leads to an unreliable communication system due to antenna gain mismatch between transceivers \cite{khan2018line,dabiri2018channel}.

\subsection{Major Contributions and Novelty}
The main contribution of this paper is to derive analytical channel models for UAV-based mmWave links while taking into account the unique characteristics of mmWave links along with the effects of UAV random vibrations and orientation fluctuations.
To this end, we assume a uniform linear array (ULA) of antennas operating at mmWave frequencies that are mounted on each UAV. We then consider three important UAV-based mmWave communication links: (i) a direct UAV-to-UAV link (called U2U link); (ii) an aerial relay link (called U2U2U link) in which source, relay, and destination are hovering UAVs; and (iii) a relay link (called G2U2G link) in which a hovering UAV connects the ground source to the ground destination. Given these three types of links, the main contributions of this work include:
\begin{itemize}
	\item We analytically derive an accurate channel model for U2U links that explicitly factors in the random vibrations of the transceivers and the unique characteristics of signal propagation at the mmWave band. We also derive a
	closed-form expression for the outage probability of a U2U mmWave link.
	\item Then, given the accurate and tractable U2U channel model that we derived, we consider an amplify-and-forward (AF) single relay for two conventional scenarios, i.e., the U2U2U link, and the G2U2G link. For two considered relaying scenarios, we derive the closed-form expressions for the end-to-end SNR at the destination as well as the outage probability.
	\item We provide simulation results to corroborate the accuracy of the derived analytical expressions and to study  the effect of antenna directivity gain on the performance of the system under different channel conditions. Simulation results show that the performance of such links is largely dependent on the random fluctuations of hovering UAVs. Moreover, the results also show that higher antenna directivity gains achieve better performance at low SNR regime. In addition, we observe that, in the high SNR regime, lower antenna directivity gains result in a more reliable communication link.
	\item Then, based on the simulation results and under different levels of instability for the UAVs, we find the optimal antenna directivity gain that allows the establishment of a reliable links in terms of achieving minimum outage probability by balancing a tradeoff between antenna beam width and directivity gain. 
\end{itemize}
The results of this paper along with the accurate analytical derivations for channel modeling can thus be applied as a benchmark for finding the optimal antenna directivity gain of UAVs under the different levels of instability without resorting to time-consuming simulations.

The rest of this paper is organized as follows. In Section II, we present our system model, while in Section III we introduce channel distribution functions and outage probability analysis of the considered links. In Section IV, numerical results are provided to verify our analytical expressions and showcase the need for antenna pattern optimization. Finally, conclusions are drawn in Section V.

%
%
\begin{table}
	\caption{Summary of our Notations} 
	\centering 
	\begin{tabular}{l l} 
		\hline\hline \\[-1.2ex]
		{\bf Parameter} & {\bf Definition} \\ [.5ex] 
		\hline\hline \\[-1.5ex]
		{\bf UAV parameters}            &  \\[.5ex] 
		\hline \\[-1.5ex]  
		--------~~~ For U2U link& ---------\\[.3ex]
		$\theta_{\textrm{tx}}$          &  Instantaneous angular deviation (IAD) of  
		\\[-.5ex] 
		& UAV Tx in $x-z$ plane 
		\\[.2ex]
		$\theta'_{\textrm{tx}}$         & Angular boresight of UAV Tx in $x-z$ plane \\[.2ex]                                                                                    
		$\sigma^2_{\textrm{tx}}$        & Variance of angular deviations (VAD) of 
		\\[-.5ex] 
		& UAV Tx in $x-z$ plane
		\\[.2ex]                                 
		$\theta_{\textrm{rx}}$          &  IAD of UAV Rx in $x-z$ plane 
		\\[.2ex]
		$\theta'_{\textrm{rx}}$         & Angular boresight of UAV Rx in $x-z$ plane \\[.2ex]                                                                                    
		$\sigma^2_{\textrm{rx}}$        & VAD of UAV Rx in $x-z$ plane
		\\[.2ex]  
		$\theta_{\textrm{ty}}$          &  IAD of UAV Tx in $y-z$ plane
		\\[.2ex]
		$\theta'_{\textrm{ty}}$         & Angular boresight of UAV Tx in $y-z$ plane \\[.2ex]                                                                                   
		$\sigma^2_{\textrm{ty}}$        & VAD of UAV Tx in $y-z$ plane 
		\\[.2ex]
		$\theta_{\textrm{ry}}$          &  IAD of UAV Rx in $y-z$ plane  
		\\[.2ex]
		$\theta'_{\textrm{ry}}$         & Angular boresight of UAV Rx in $y-z$ plane \\[.2ex]                                                                                    
		$\sigma^2_{\textrm{ry}}$        & VAD of UAV Rx in $y-z$ plane
		\\[1ex]  
		-----~~~ For relaying link& ---------\\[.3ex]
		$\theta_{s}$ and $\theta_{d}$         &  IAD of source and destination
		\\[.2ex]
		$\theta'_{s}$  and $\theta'_{d}$       & Angular boresight of source and destination \\[.2ex]                                                                                    
		$\sigma^2_{s}$ and $\sigma^2_{s}$       & VAD of source and destination
		\\[.2ex]  
		$\theta_{\textrm{rs}}$          &  IAD of relay antenna tilted towards source 
		\\[.2ex]
		$\theta_{\textrm{rd}}$          &  IAD of relay antenna tilted towards destination 
		\\[.2ex]
		$\theta_{R}$           &  $=\theta_{\textrm{rs}}=\theta_{\textrm{rd}}$ 
		\\[.2ex]
		$\theta'_{R}$         & Angular boresight of relay \\[.2ex]                                                                                    
		$\sigma^2_{R}$        & VAD of relay
		\\[.2ex]  
		\hline \\[-1.5ex]
		{\bf Antenna parameters}        &  \\[.5ex] 
		\hline \\[-1.5ex]  
		$f_c$                           &  Carrier frequency
		\\[.3ex]
		$\lambda$                       & Wavelength \\[1ex]                                                                                    
		$\sigma^2$                      & Normalized thermal noise
		\\[.3ex]  
		$N$                             &  Number of elements of ULA antenna
		\\[.3ex]
		$M$                             & Number of sectors (used for approximating
		\\[-.5ex] 
		&  of main-lobe)
		\\[.3ex] 
		--------~~~ For U2U link& ---------\\[.3ex]
		$\mathbb{G}\left(\theta_{\textrm{ty}},\theta_{\textrm{ry}}\right)$          
		&  Instantaneous directivity gain
		\\[.3ex]
		$G_t(\theta_{\textrm{ty}})$ and  $G_r(\theta_{\textrm{ry}})$   & Array gain of Tx and Rx 
		\\[.3ex]                                                                                    
		-----~~~ For relaying link& ---------\\[.3ex]
		$G_{r1}(\theta_{\textrm{rs}})$  & relay array gain tilted towards source
		\\[.3ex] 
		$G_{r2}(\theta_{\textrm{rd}})$  & relay array gain tilted towards destination
		\\[.3ex] 
		$G_s(\theta_s)$     & Array gain of source 
		\\[.3ex]  
		$G_s(\theta_d)$     & Array gain of destination
		\\[.3ex]
		$\mathbb{G}_{\textrm{sr}}(\theta_s,\theta_{\textrm{rs}})$  &  directivity gain of source-to-relay  
		\\[.3ex]  
		$\mathbb{G}_{\textrm{dr}}(\theta_d,\theta_{\textrm{rd}})$  &  directivity gain of relay-to-destination   
		\\[.3ex]                                                                                                 
		%
		%
		%
		%
		%
		%
		%
		%
		%
		%
		\hline \\[-1.5ex]
		{\bf Channel parameters}        &  \\[.5ex] 
		\hline \\[-1.5ex]  
		$\alpha$                        & Small-scale channel coefficient
		\\[.3ex]
		$h_L(Z)$                        & Large-scale channel coefficient \\[1ex]                                                                                    
		$h_b$                           & Average of building height of cities
		\\[.3ex]  
		$\zeta$                         &  $=\alpha^2$
		\\[.3ex]
		$m$                             & Nakagami fading parameter \\[1ex]                                                                                    
		$\gamma$                        & Instantaneous SNR at the Rx
		\\[.3ex]  
		$\gamma_{\textrm{th}}$          &  SNR threshold
		\\[.1ex]
		$\Sigma I$                      &  $=\Sigma I_{\textrm{t}}+\Sigma I_{\textrm{other}}$, ICI
		\\[.3ex]
		$\Sigma I_{\textrm{t}}$         &  ICI due to Doppler shift 
		\\[.3ex]
		$\Sigma I_{\textrm{other}}$     &  ICI due to other transmitters
		\\[.3ex]
		$f_{\textrm{dop}}$              &  Doppler frequency shift
		\\[.5ex]
		%
		%
		%
		\hline \\[-1.5ex]
		{\bf Functions}                 &  \\[.5ex] 
		\hline \\[-1.5ex]  
		$\Gamma(\cdot)$                     &  Gamma function
		\\[.3ex]
		$\Gamma(\cdot,\cdot)$                   & Incomplete Gamma function  \\[1ex]                                                                                    
		$G_{m,n}^{p,q}$                 & Meijer's G-function    
		\\[.3ex]
		$Q(\cdot)$                          & {\it Q}-function \\[1ex]                                                                                    
		$\delta(\cdot)$                     & Delta function
		\\[.5ex]
		\hline\hline 
	\end{tabular}
	\label{tabs} 
\end{table}
%
%

%
\begin{figure}
	\centering
	\subfloat[] {\includegraphics[width=3.3 in]{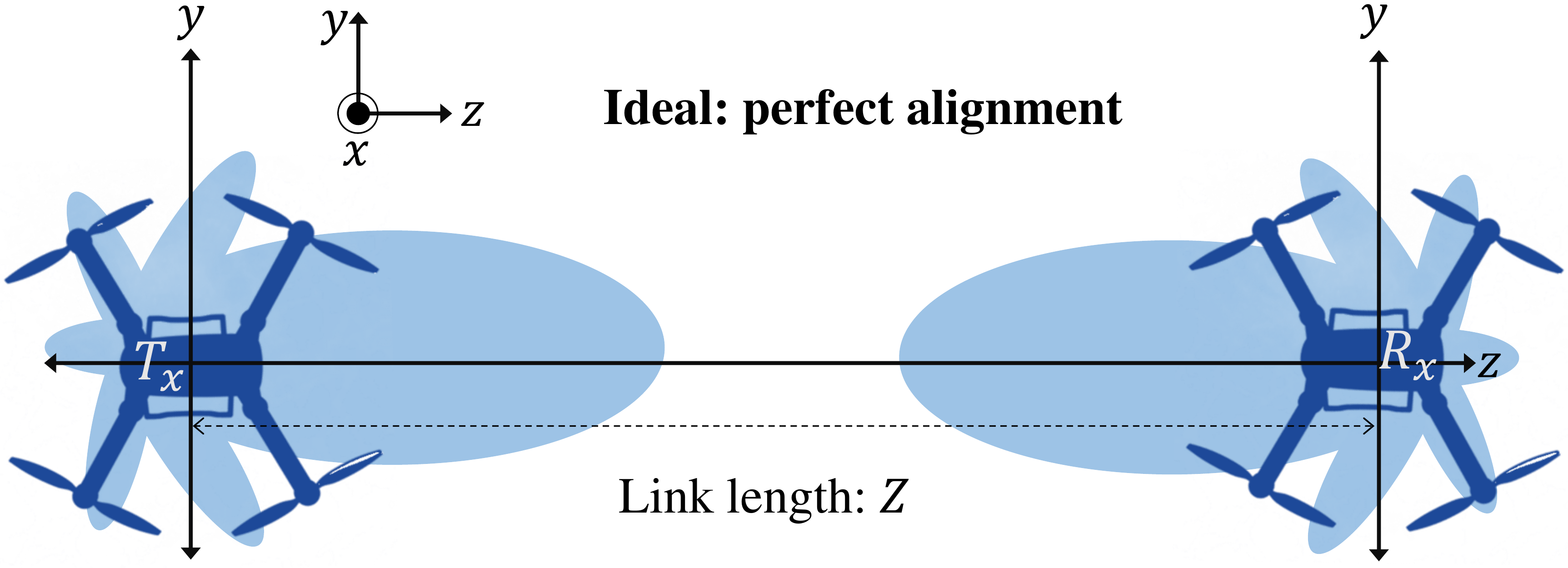}
		\label{UAV_1D}
	}
	\hfill
	\subfloat[] {\includegraphics[width=3.3 in]{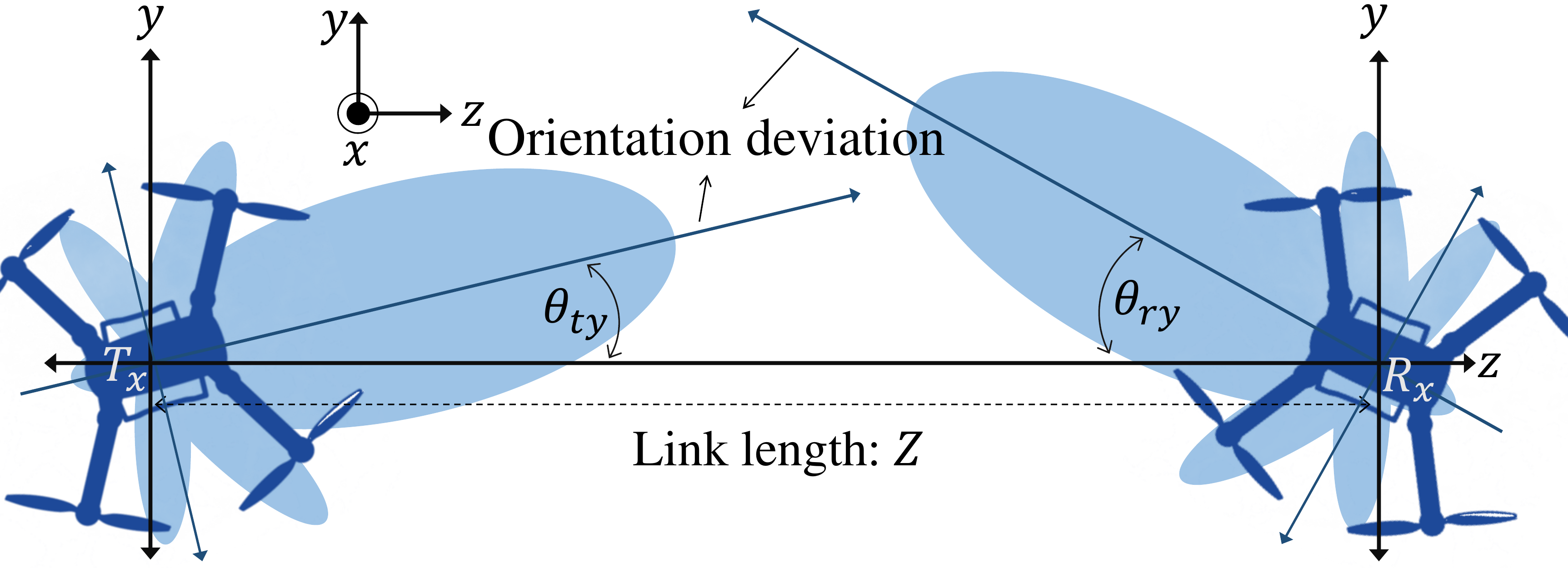}
		\label{UAV_2D}
	}
	\hfill
	\subfloat[] {\includegraphics[width=3.3 in]{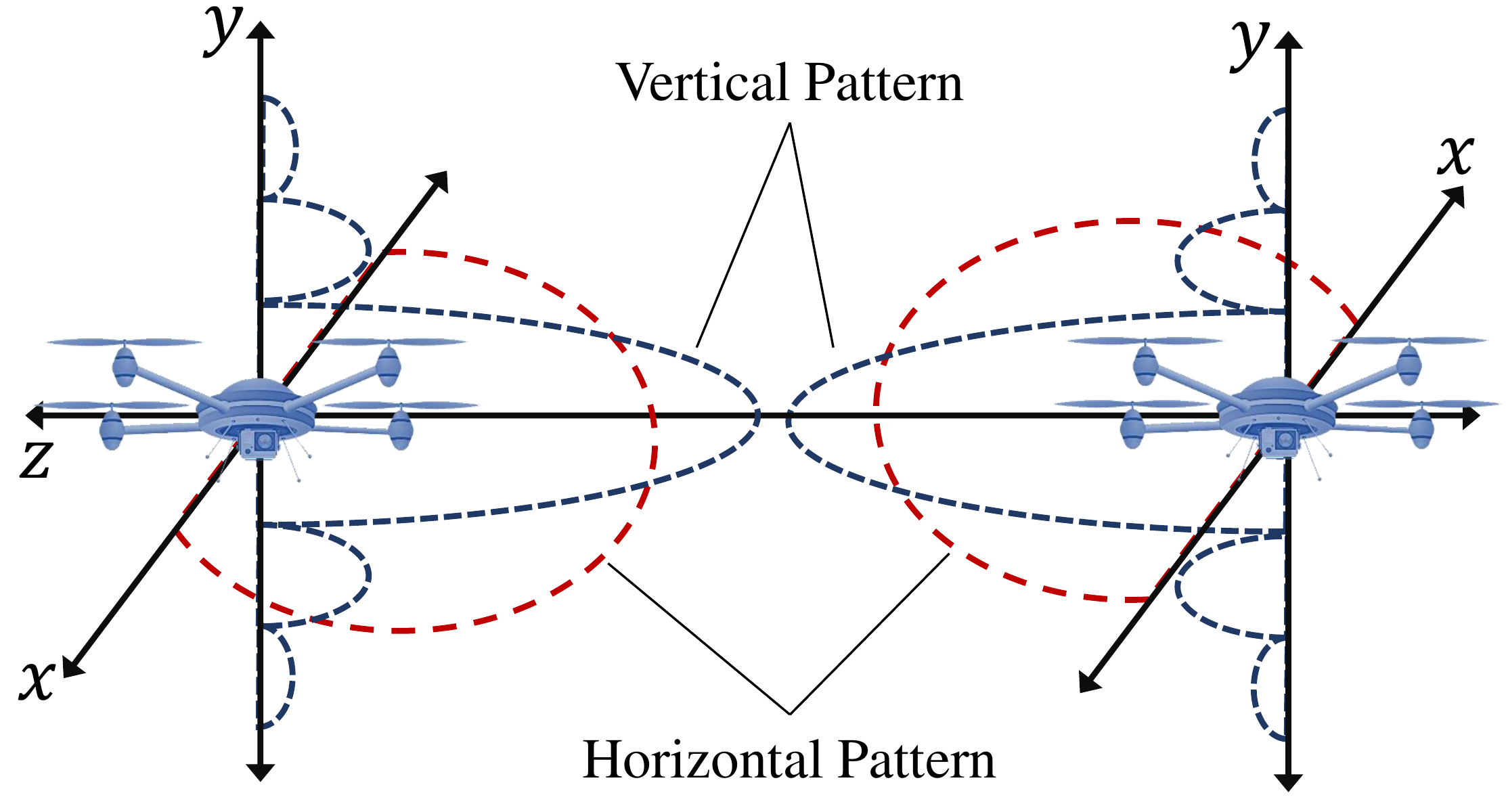}
		\label{UAV_3D}
	}
	\caption{U2U system consisting of two hovering UAV nodes equipped with directional beam a) 2-D configuration when UAVs are perfectly aligned (ideal assumption), b) 2-D configuration of orientation deviation due to hovering fluctuations of UAVs, and c) 3-D configuration {in which the directional vertical pattern is placed on $y-z$ plane, and approximately flat horizontal pattern is placed on $x-z$ plane.} }
	\label{uav}
\end{figure}
%
\section{System Model}

Consider a UAV-based mmWave communication link that consists of two hovering UAVs and are used to provide a high bandwidth point-to-point link  as shown in Fig. \ref{uav}. We assume that UAVs are hovering at a distance of $Z$ from each other and are capable of performing directional beamforming by using mmWave antenna arrays to transmit the signal array patterns in the direction of $z$-axis. Also, the mean of the UAVs' positions are respectively located at $[0, 0, 0]$ and $[0, 0, Z]$ in a Cartesian coordinate system $[x,y,z]\in \mathbb{R}^{1\times3}$, and are known at the transceiver (e.g., shared by transmitting  periodic broadcast messages). The considered UAVs will use mechanical and control systems with pinpoint accuracy  and, thus, the degree of instability of the hovering UAVs is in the order of several $ \rm mrad$ \cite{orsag2017dexterous}. In practice, as shown in Fig. \ref{UAV_2D}, the instantaneous orientations of the UAVs can randomly fluctuate away from their means which leads to deviations in the angle-of-departure (AoD) and angle-of-arrival (AoA) of the transmission pattern. At the transmitter side, the AoD deviations are given by $\theta_{\textrm{tx}}$ and $\theta_{\textrm{ty}}$ in the $x-z$ plane and the $y-z$ plane, respectively. At the receiver side, the AoA deviations are given by $\theta_{\textrm{rx}}$ and $\theta_{\textrm{ry}}$ in the $x-z$ plane and the $y-z$ plane, respectively. Moreover, based on the central limit theorem, the deviations of the UAVs' orientations are considered to be Gaussian distributed \cite{dabiri2018channel,safi2019spatial}. Therefore, we have $\theta_i\sim \mathcal{N}(\theta'_i,\sigma_i^2)$, where $\theta'_i$ is the boresight direction of the antennas and $i\in\{\textrm{tx, ty, rx, ry}\}$. A summary of our notations is provided in Table \ref{tabs}.

The transmitted pattern can be decomposed into vertical and horizontal patterns. To transmit a directional vertical pattern, a ULA consisting of $N$ antenna element operating at mmWave frequencies is employed in which the antenna elements are uniformly arranged in a single line with spacing of $\lambda/2$. For this setup, the transmitted horizontal pattern can be assumed to be approximately constant.
In Fig. \ref{UAV_3D}, a graphical representation of the considered three-dimensional radiation patterns is shown where vertical and horizontal antenna patterns are in the $y-z$ plane and the $x-z$ plane, respectively. Since the antenna gain in the $x-z$ plane is approximately constant, one can reasonably neglect the effect of the orientation deviations of the aerial nodes in the $x-z$ plane on the performance of considered system. However, due to the directional pattern shape which requires a careful alignment between transceivers, the performance of the considered system is highly dependent on the orientation deviations of the hovering UAVs in the $y-z$ plane.

Given these assumptions, the instantaneous SINR at the receiver side will be:
\begin{align}
\label{snrr1}
\gamma(\alpha,\theta_{\textrm{ty}},\theta_{\textrm{ry}})  = \frac{|\alpha|^2 h_L(Z) G(\theta_{\textrm{ty}},\theta_{\textrm{ry}})}
{ \Sigma I + \sigma^2},
\end{align}
where $\sigma^2$ is the normalized thermal noise power, $\alpha$ is the small scale fading coefficient, and $h_L(Z)$ is the path loss coefficient.
Moreover in (\ref{snrr1}), $\Sigma I=\Sigma I_{\textrm{d}}+\Sigma I_{\textrm{r}}$, where $\Sigma I_{\textrm{d}}$ and $\Sigma I_{\textrm{r}}$ are the inter-carrier interference due to Doppler spread, and radio interference due to the other transmitters, respectively.
Note that, by using high directional radio patterns at the receiver, $\Sigma I_{\textrm{r}}$ can be effectively eliminated \cite{lin2018sky,3gpp1}.
Also, $\Sigma I_{\textrm{t}}$ is caused by Doppler spread and it is proportional to $\Sigma I_{\textrm{t}} \propto \left[1-\textrm{sinc}^2(f_{\textrm{d}}T_s)\right]$, where $T_s$ is the symbol duration, $f_{\textrm{d}} = \frac{f_c \nu}{c}$ is the Doppler frequency shift, $c=3\times 10^8$ (in m/s) is the speed of light, $\nu$ (in m/s) is the relative moving velocity, and $f_c$ (in GHz) is the carrier frequency \cite{sesia2011lte}.
Moreover, in \cite{zhou2019beam}, it was shown that for a moving UAV with $\nu\leq 10$\,m/s,  the impact of the Doppler spread is negligible. In our setup, we assume that UAVs are hovering at a fixed position (i.e., multi-rotor UAVs or tethered balloons) and there is no relative velocity between communication nodes; therefore, there will be no Doppler spreading effect. As a result, \eqref{snrr1} can be simplified as
\begin{align}
\label{snr1}
\gamma(\alpha,\theta_{\textrm{ty}},\theta_{\textrm{ry}})  = \frac{|\alpha|^2 h_L(Z) G(\theta_{\textrm{ty}},\theta_{\textrm{ry}})}
{  \sigma^2}.
\end{align}

Since no standardized results for UAV-based communications at mmWave bands exist, we consider the recent 3GPP report to set the path loss parameters  \cite{3gppf}. These parameters are valid for a BS height up to  150 m and are expressed as follows:
\begin{align}
\label{loss}
h_{\textrm{L,dB}}(Z) &= 20 \log_{10}\left(\frac{40 \pi Z f_c}{3}\right)  \\
&+ \min\left\{0.03 h_b ^{1.73},10\right\}\times \log_{10}(Z) \nonumber \\
& - \min\left\{0.044 h_b ^{1.73},14.77\right\}
+0.002\, Z \log_{10}(h_b),  \nonumber
\end{align}
where  $h_b$ (in m) is the average of building height of city.

Also in (\ref{snr1}), $G(\theta_{\textrm{ty}},\theta_{\textrm{ry}})$ is the instantaneous directivity gain given by \cite{yu2017coverage}:
\begin{align}
\label{G_direct}
\mathbb{G}(\theta_{\textrm{ty}},\theta_{\textrm{ry}}) =
\underbrace{
\frac{\sin^2(\pi N \theta_{\textrm{ty}})}{N \sin^2(\pi \theta_{\textrm{ty}} )}
}_{G_t(\theta_{\textrm{ty}})}
\underbrace{
\frac{\sin^2(\pi N \theta_{\textrm{ry}})}{N \sin^2(\pi \theta_{\textrm{ry}} )}
}_{ G_r(\theta_{\textrm{ry}})},
\end{align}
where $ G_t(\theta_{\textrm{ty}})$ and $ G_r(\theta_{\textrm{ry}}) $ are the actual array gains of the transmitter and receiver, respectively.
Moreover, from the measurement results provided in \cite{goddemeier2015investigation}, for a low altitude communication link between UAVs, Rician and Nakagami distributions were shown to be highly promising models  that can be mathematically fitted into the experimentally measured data.
Since the Nakagami distribution is a universal model that can capture various channel conditions, we apply it to model small-scale fading.
Let $\alpha$ be a Nakagami random variable (RV), and, hence, $\zeta= \alpha^2$ will be a normalized Gamma RV given by:
\begin{align}
\label{Gamma}
f_\zeta (\zeta) = \frac{m^m  \zeta^{m-1}}{\Gamma(m)}     \exp(-m \zeta), ~~~\zeta>0,
\end{align}
\textcolor{black}{where $m$ is the Nakagami fading parameter and $\Gamma(\cdot)$ is the Gamma function \cite{goddemeier2015investigation}.}

 In practice, a highly directional beam is used to compensate the high free-space path loss at the mmWave band.  Hence, in addition to the channel fading, fluctuations in the orientation of the UAVs (due to the effect of wind, mechanical and control system flaws, antenna and BS payload, etc.) can  lead to beam misalignment and adversely affect the link performance and channel capacity. To capture these effects, we define the outage capacity, i.e., the probability with which the instantaneous capacity falls bellow a certain threshold $\mathcal{C}_{\textrm{th}}$, as the figure of merit to determine the reliability of the considered UAV-based communication system. The outage capacity can be defined as follows:
\begin{align}
\label{xd}
\mathbb{C}_{\textrm{out}} &= {\textrm{Pr}}\{\log_2(1+\gamma)<\mathcal{C}_{\textrm{th}}\}
= F_{\gamma}(\gamma_{\textrm{th}}),
\end{align}
where $F_x(x)$ is the cumulative distribution function (CDF) of RV $x$, $\gamma$ is the instantaneous SNR, and $\gamma_{\textrm{th}}=2^{\mathcal{C}_{\textrm{th}}}-1$ is the SNR threshold.
%
\begin{figure}
	\begin{center}
		\includegraphics[width=3.2 in]{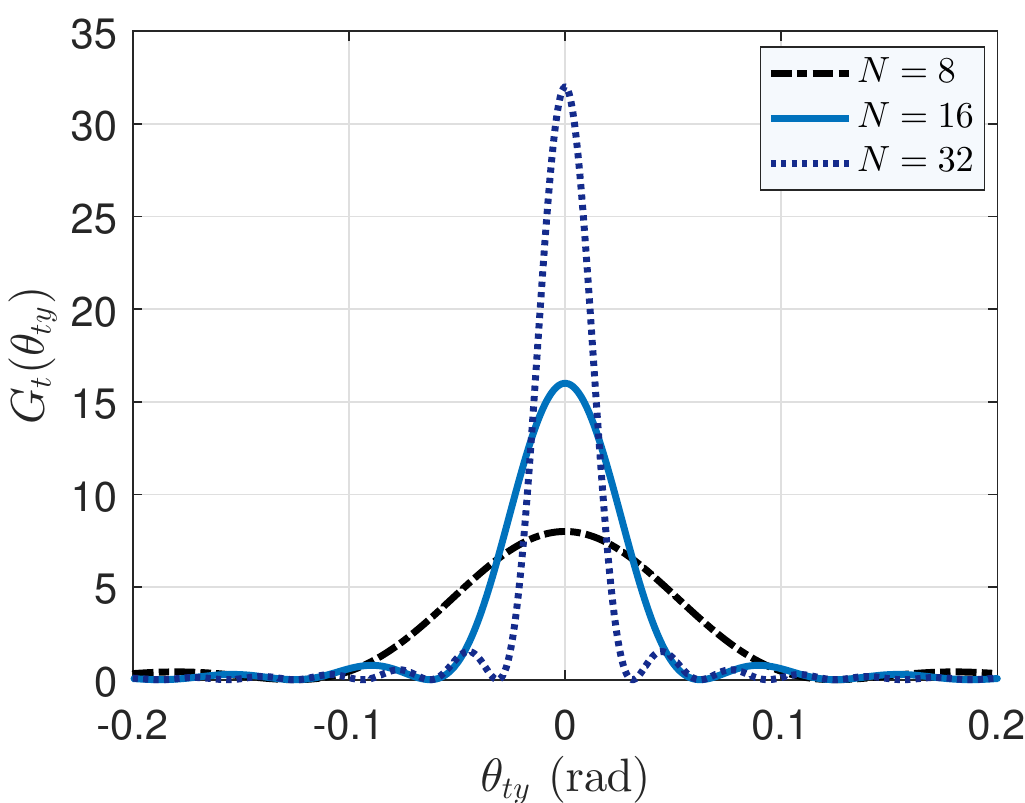}
		\caption{Actual array gain for different values of $N$.}
		\label{M1}
	\end{center}
\end{figure}
%
%
\begin{figure}
	\begin{center}
		\includegraphics[width=3.2 in]{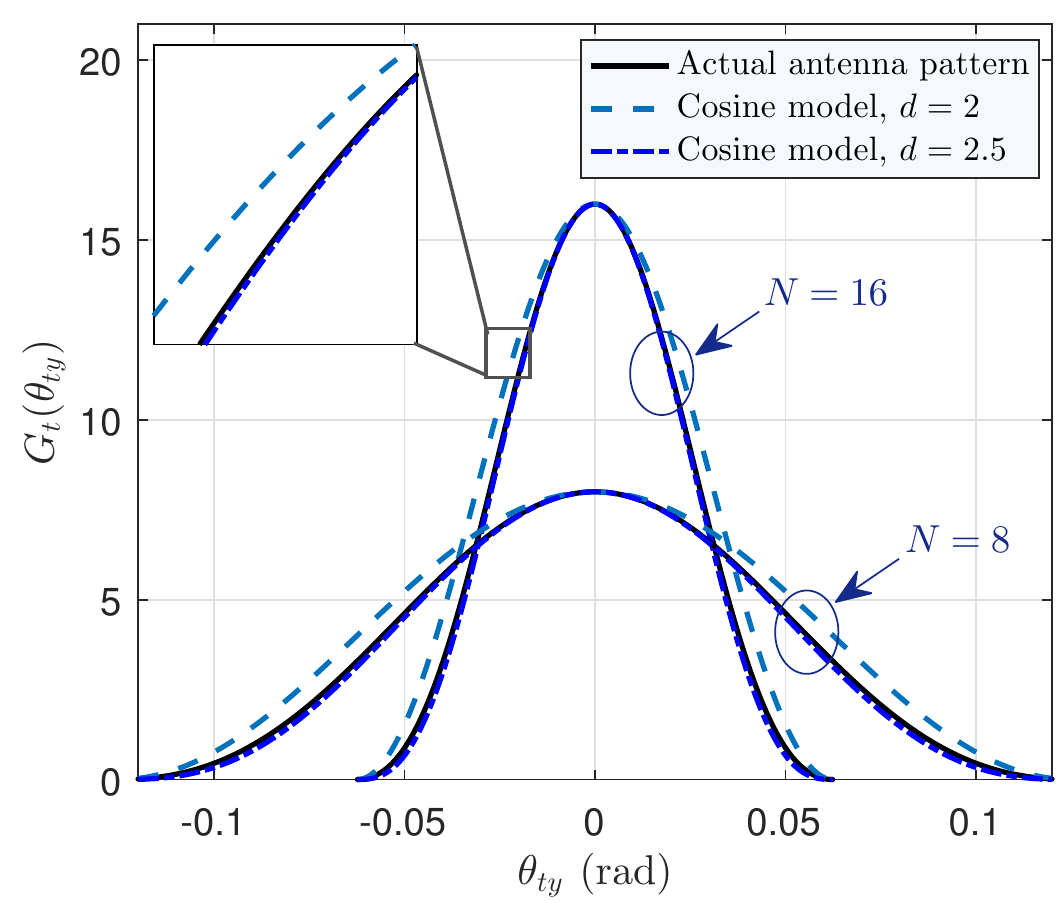}
		\caption{{Comparison between the values of $G_{\textrm{ty}}$ in \eqref{gap} for $d=2$ and $d=2.5$ with the actual array gain of \eqref{G_direct}.}}
		\label{M2}
	\end{center}
\end{figure}
%

{From \eqref{snr1}, it can observe that the instantaneous SNR is composed of the deterministic loss parameter $h_L$, and several random variables, i.e., the small-scale fading coefficient $\alpha$, the AoD deviations $\theta_{\textrm{ty}}$, and the AoA deviations $\theta_{\textrm{ry}}$. In the next section, for the considered UAV-based mmWave communication scenarios, our aim is to derive the closed-form expressions of the SNR distribution at the receiver  which can be used for easily evaluating the performance of hovering UAV-based mmWave links without performing time-consuming simulations.}

\section{Channel Distribution Function}
Next, we first develop a channel model for the U2U link. Then, we proceed to determine the end-to-end SNR for a network consisting of an aerial relay node (i.e., U2U2U \& G2U2G links).
\subsection{U2U Link}
Here, we derive a closed-form expression for the probability density function (PDF) for the SNR of the considered U2U link.
In Fig. \ref{M1}, we show the actual array gain versus $\theta_{\textrm{ty}}$ for different values of $N$. Clearly, for large values of $N$, it can be readily observed that the main lobe peak-gain is much larger than the side lobe peak-gain. Therefore, for a point-to-point link, it is reasonable to approximate the actual antenna array gain by only its main lobe. According to \cite[ (9)]{yu2017coverage}, $G_t(\theta_{\textrm{ty}}) $ (or equally $G_r(\theta_{\textrm{ry}})$) can be approximated by the following model:
\begin{align}
\label{gap}
G_{\textrm{ty}}(\theta_{\textrm{ty}}) \simeq \left\{
\begin{array}{rl}
N\cos\left(\frac{\pi N}{2}\theta_{\textrm{ty}} \right)^d,& ~~~~ \theta_{\textrm{ty}}< \frac{1}{N},\\
0,~~~~~~~~~~~~~~~~~& ~~~~ {\textrm{otherwise.}}
\end{array} \right.
\end{align}
In \cite{yu2017coverage}, for analytical tractability, the value of $d$ is set to 2. However, from Fig. \ref{M2},  $d=2.5$ gives a more accurate approximation of actual array gain and, hence, we set $d=2.5$. Now, by sectorizing \eqref{gap}, we propose a model called the \emph{sectorized-cosine} model given by
\begin{align}
\label{sect}
G_{\textrm{ty}}(\theta_{\textrm{ty}},M) \simeq& \left\{
\begin{array}{rl}
N \cos\left(\frac{\pi N i}{2M N} \right)^{2.5},& ~~~ \frac{i}{MN} \leq|\theta_{\textrm{ty}}|< \frac{i+1}{MN}, \\
0,~~~~~~~~~~~~~~~~~& ~~~ {\rm otherwise,} \\
\end{array} \right. \nonumber \\
\end{align}
where $ i\in\{0,1,...,M-1\}$. {Figure \ref{M3}  plots the sectorized-cosine model versus $\theta_{\textrm{ty}}$  for $M=5$ and $10$.} Obviously, the accuracy of the proposed model increases by increasing $M$ at the cost of more complexity. Hence, choosing an optimal value for $M$ involves a tradeoff between tolerable complexity and desirable accuracy. This tradeoff is thoroughly studied in Section IV. \\

{\bf Theorem 1.}
{\it
For considered UAV-based mmWave communication links, the closed-form expressions for the PDF and CDF of the instantaneous SNR at the receiver are derived respectively as:
\begin{align}
\label{po1}
f_\gamma(\gamma) = \sum_{i=0}^{M-1} \sum_{j=0}^{M-1}
B'_{ij}
\gamma^{m-1} \exp\left(-\frac{m \sigma^2\, \gamma}{h_L(Z)B_{ij}(M,N) }\right),
\end{align}
and
\begin{align}
\label{pos1}
F_\gamma(\gamma) =& 2\sum_{i=0}^{M-1} \sum_{j=0}^{M-1}
B'_{ij}
\left( \frac{h_L(Z)B_{ij}(M,N) }{m \sigma^2} \right)^{m} \nonumber \\
&\times \mathbb{V}\left( m, \frac{m \sigma^2\, \gamma}{h_L(Z)B_{ij}(M,N) }  \right),
\end{align}
where
\begin{align}
\label{cd3}
&B_{ij}(M,N) =
N^2 \cos\left(\frac{\pi N i}{2M N} \right)^{2.5}  \cos\left(\frac{\pi N j}{2M N} \right)^{2.5}, \\
\label{p1}
&B'_{ij} = \frac{(\sigma^2\,m)^m }{\Gamma(m)}
\frac{A_{\textrm{r}j}\left(\theta'_{\textrm{ry}},\sigma_{\textrm{ry}}\right)A_{\textrm{t}i}\left(\theta'_{\textrm{ty}},\sigma_{\textrm{ty}}\right)}
{\left(h_L(Z) B_{ij}(M,N) \right)^m   },\\
\label{a}
&A_{\textrm{t}i}\left(\theta'_{\textrm{ty}},\sigma_{\textrm{ty}}\right) =
Q\left(\frac{i+ N M \theta'_{\textrm{ty}} }{N M \sigma_{\textrm{ty}}} \right)
-  Q\left(\frac{i+1+ N M \theta'_{\textrm{ty}} }{N M \sigma_{\textrm{ty}}} \right) \nonumber \\
&~~~~~~~~~+Q\left(\frac{i- N M \theta'_{\textrm{ty}} }{N M \sigma_{\textrm{ty}}} \right)
-  Q\left(\frac{i+1- N M \theta'_{\textrm{ty}} }{N M \sigma_{\textrm{ty}}} \right),\\
\label{ab}
&A_{\textrm{r}j}\left(\theta'_{\textrm{ry}},\sigma_{\textrm{ry}}\right) =
Q\left(\frac{j+ N M \theta'_{\textrm{ry}} }{N M \sigma_{\textrm{ry}}} \right)
-  Q\left(\frac{j+1+ N M \theta'_{\textrm{ry}} }{N M \sigma_{\textrm{ry}}} \right) \nonumber \\
&~~~~~~~~~+Q\left(\frac{j- N M \theta'_{\textrm{ry}} }{N M \sigma_{\textrm{ry}}} \right)
-  Q\left(\frac{j+1- N M \theta'_{\textrm{ry}} }{N M \sigma_{\textrm{ry}}} \right),
\end{align}
and $Q(\cdot)$ is the {\it Q}-function, and $\mathbb{V}(\cdot,\cdot)$ is the incomplete Gamma function \cite[(8.350.1)]{jeffrey2007table}.
%
%
%
}
\begin{IEEEproof}
To present \eqref{sect} in a more tractable manner, we can rewrite it as follows:
\begin{align}
\label{kgq}
 G_{\textrm{ty}}(\theta_{\textrm{ty}},M) \simeq&
N  \Pi\left( M N \theta_{\textrm{ty}}  \right)
+\sum_{i=1}^{M-1}
N \cos\left(\frac{\pi N i}{2M N} \right)^{2.5}  \\
&\times \left[ \Pi\left( \frac{M N |\theta_{\textrm{ty}}|}{i+1}  \right)  -  \Pi\left( \frac{M N |\theta_{\textrm{ty}}|}{i}  \right)  \right], \nonumber
\end{align}
where
$
\Pi(x)= \left\{
\begin{array}{rl}
1& ~~~ {\rm for}~~~ |x|\leq 1 \\
0& ~~~ {\rm for}~~~ |x|> 1 \\
\end{array} \right. .
$
%
\begin{figure}
	\begin{center}
		\includegraphics[width=3.2 in]{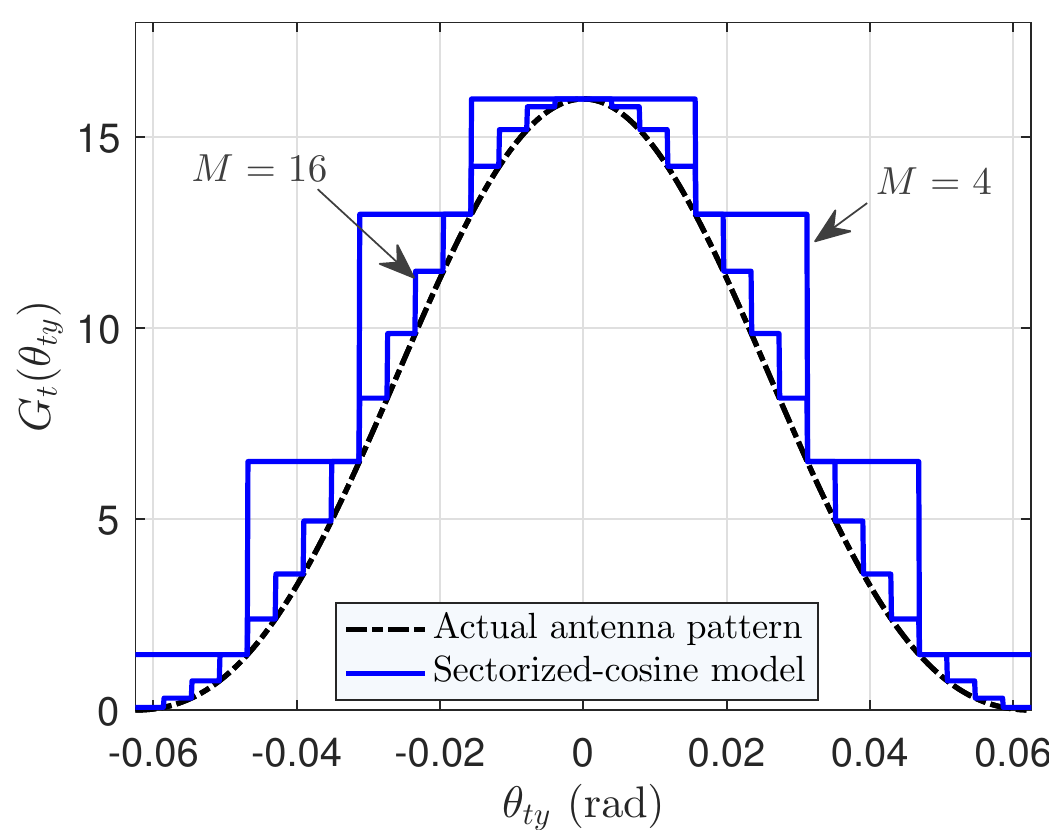}
		\caption{Sectorized-cosine model for $M=4$ and $16$ and comparison with actual antenna pattern gain for $N=16$.}
		\label{M3}
	\end{center}
\end{figure}
{As mentioned in Section II,  $\theta_m $ denotes the instantaneous orientations of UAVs (or equivalently the instantaneous orientations of antennas mounted on UAVs) and follows a Gaussian distribution   $\theta_m\sim \mathcal{N}(\theta'_m,\sigma_m^2)$ where $m\in\{\textrm{ty, ry}\}$.} Hence, the PDF of $G_{\textrm{ty}}(\theta_{\textrm{ty}},M)$ will be given by:
\begin{align}
\label{cd1}
f_{G_{\textrm{ty}}}(G_{\textrm{ty}}) =& \sum_{i=0}^{M-1} A_{\textrm{t}i}\left(\theta'_{\textrm{ty}},\sigma_{\textrm{ty}}\right)  \\
 &\times \delta\left(G_{\textrm{ty}} - N \cos\left(\frac{\pi N i}{2M N} \right)^{2.5} \right),  \nonumber
\end{align}
where
\begin{align}
\label{aoo}
&A_{\textrm{t}i}\left(\theta'_{\textrm{ty}},\sigma_{\textrm{ty}}\right) =
Q\left(\frac{i+ N M \theta'_{\textrm{ty}} }{N M \sigma_{\textrm{ty}}} \right)
-  Q\left(\frac{i+1+ N M \theta'_{\textrm{ty}} }{N M \sigma_{\textrm{ty}}} \right) \nonumber \\
&~~~~~~~~~+Q\left(\frac{i- N M \theta'_{\textrm{ty}} }{N M \sigma_{\textrm{ty}}} \right)
-  Q\left(\frac{i+1- N M \theta'_{\textrm{ty}} }{N M \sigma_{\textrm{ty}}} \right),
\end{align}
and $\delta(\cdot)$ is the Dirac delta function.
Note that, the PDF of RV $G_{\textrm{ry}}(\theta_{\textrm{ry}},M)$ can be obtained similar to \eqref{cd1} by replacing subscript ty with ry.  From \eqref{G_direct} and \eqref{cd1}, the PDF of directivity gain $\mathbb{G}(\theta_{\textrm{ty}},\theta_{\textrm{ry}})$ conditioned on the transmitter array gain can be given as:
\begin{align}
\label{sc}
f_{\mathbb{G}|G_{\textrm{ty}}}(\mathbb{G})
=& \sum_{i=0}^{M-1} \frac{A_{\textrm{r}i}\left(\theta'_{\textrm{ry}},\sigma_{\textrm{ry}}\right)}{G_{\textrm{ty}}} \\
&\times \delta\left(\frac{\mathbb{G}}{G_{\textrm{ty}}} - N \cos\left(\frac{\pi N i}{2M N} \right)^{2.5} \right).\nonumber
\end{align}
By using \eqref{cd1} and \eqref{sc}, the closed-form expression for the PDF of directivity gain $\mathbb{G}$ is derived as
\begin{align}
\label{cd2}
f_\mathbb{G}(\mathbb{G}) =& \int f_{\mathbb{G}|G_{\textrm{ty}}}(\mathbb{G}) f_{G_{\textrm{ty}}}(G_{\textrm{ty}}) dG_{\textrm{ty}}    \\
=& \sum_{i=0}^{M-1} \sum_{j=0}^{M-1} \int
\frac{A_{\textrm{r}j}\left(\theta'_{\textrm{ry}},\sigma_{\textrm{ry}}\right)A_{\textrm{t}i}\left(\theta'_{\textrm{ty}},\sigma_{\textrm{ty}}\right)}{G_{\textrm{ty}}} \nonumber \\
&\times \delta\left(G_{\textrm{ty}} - N \cos\left(\frac{\pi N i}{2M N} \right)^{2.5} \right) \nonumber \\
&\times  \delta\left(\frac{\mathbb{G}}{G_{\textrm{ty}}} - N \cos\left(\frac{\pi N j}{2M N} \right)^{2.5} \right) dG_{\textrm{ty}} \nonumber \\
=& \sum_{i=0}^{M-1} \sum_{j=0}^{M-1}
\frac{A_{\textrm{r}j}\left(\theta'_{\textrm{ry}},\sigma_{\textrm{ry}}\right)A_{\textrm{t}i}\left(\theta'_{\textrm{ty}},\sigma_{\textrm{ty}}\right)}
{B_{ij}(M,N)   }   \nonumber \\
&\times \delta\left( \mathbb{G} -  B_{ij}(M,N) \right), \nonumber
\end{align}
where
\begin{align}
\label{cd300}
B_{ij}(M,N) =
N^2 \cos\left(\frac{\pi N i}{2M N} \right)^{2.5}  \cos\left(\frac{\pi N j}{2M N} \right)^{2.5}.
\end{align}
Finally, from \eqref{snr1}, \eqref{Gamma} and \eqref{cd2}, the PDF of RV $\gamma$ is obtained from
\begin{align}
\label{pooo1}
f_\gamma(\gamma) =& \int_0^\infty \frac{\sigma^2}{h_L(Z) \zeta}f_\mathbb{G}\left( \frac{\sigma^2\gamma}{h_L(Z) \zeta} \right)
f_\zeta(\zeta) d\zeta  \\
=& \sum_{i=0}^{M-1} \sum_{j=0}^{M-1}
B'_{ij}
\gamma^{m-1} \exp\left(-\frac{m \sigma^2\, \gamma}{h_L(Z)B_{ij}(M,N) }\right),\nonumber
\end{align}
where
\begin{align}
\label{poo1}
B'_{ij} = \frac{(\sigma^2\,m)^m }{\Gamma(m)}
\frac{A_{\textrm{r}j}\left(\theta'_{\textrm{ry}},\sigma_{\textrm{ry}}\right)A_{\textrm{t}i}\left(\theta'_{\textrm{ty}},\sigma_{\textrm{ty}}\right)}
{\left(h_L(Z) B_{ij}(M,N) \right)^m   }.
\end{align}
Moreover, from \cite[(3.381.1)]{jeffrey2007table}, the CDF of RV $\gamma$ can be obtained as \eqref{pos1}.
\end{IEEEproof}

We note that the accuracy of the derived analytical expressions depends on the variable $M$ that is used for the approximation of antenna pattern. By increasing $M$, the accuracy improves at the expense of more complexity. Hence, the optimal value of $M$ should  satisfy a predefined accuracy as well as a tolerable complexity. As seen later, $M=20$ is a good choice and achieves the analytical results close to the simulation results. Also from \eqref{po1} and \eqref{pos1}, the system performance metrics for an U2U link, e.g., channel capacity, outage probability, and bit error rate can be analytically
developed without resorting to time-consuming simulations.

%
\begin{figure}
	\centering
	\subfloat[] {\includegraphics[width=3.3 in]{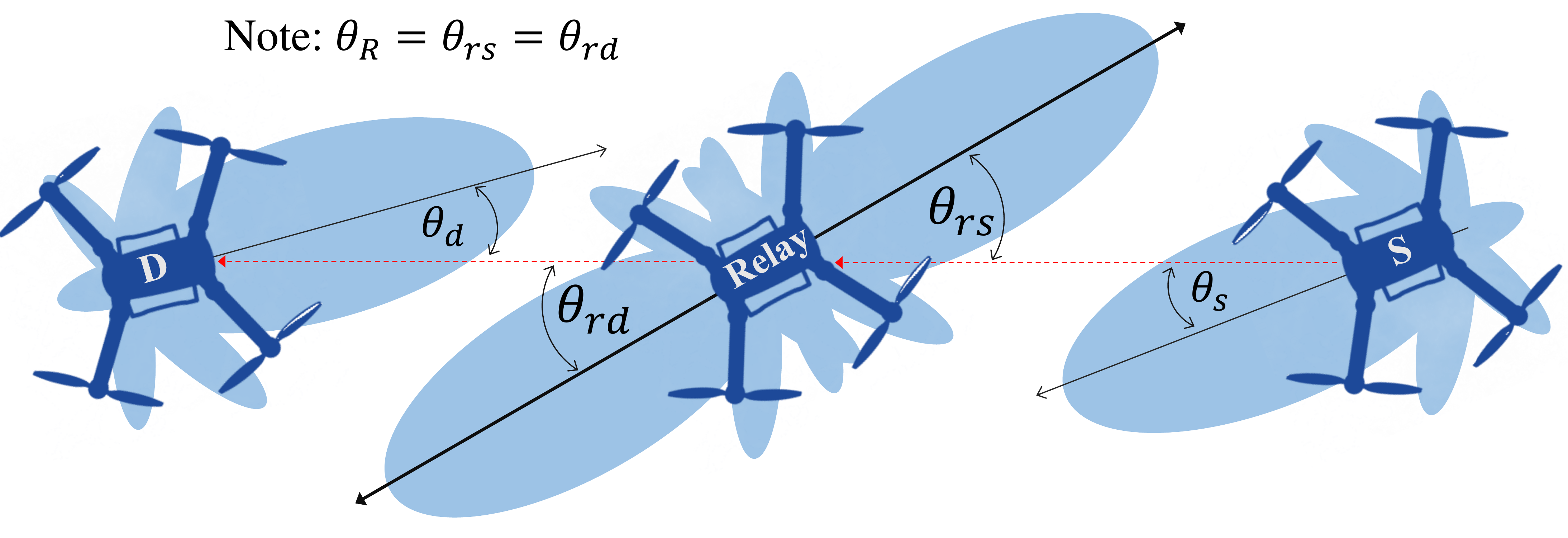}
		\label{relay1}
	}
	\hfill
	\subfloat[] {\includegraphics[width=3.3 in]{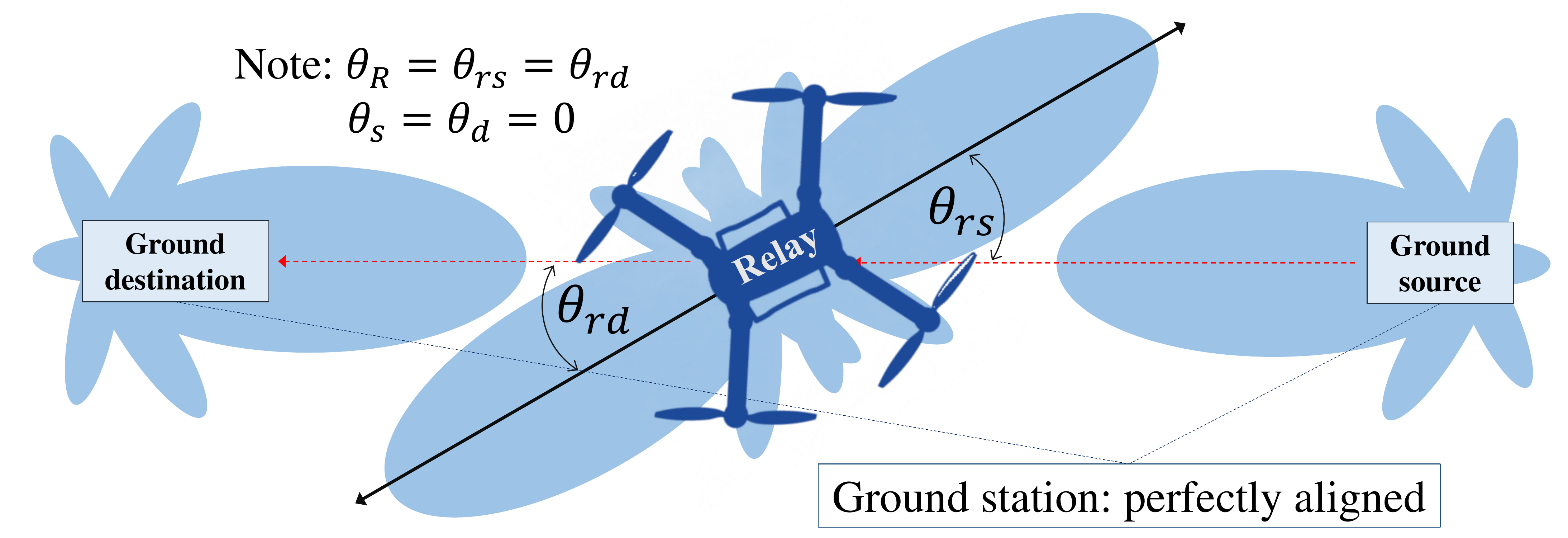}
		\label{relay2}
	}
	\caption{ 2-D configuration of UAV-based relaying systems which are connected by directional beams for a) U2U2U link and b)  G2U2G link.}
	\label{relay}
\end{figure}



%
%
\subsection{U2U2U Link}
Next, we consider an aerial relay-assisted communication system in which an aerial node relays transmitted signal from the aerial source node to the aerial destination node.
The relay node has two antennas: 1) a directional receive antenna tilted towards the source node with the antenna array gain $G_{r1}$; and 2) a directional transmit antenna tilted towards the destination node with the antenna array gain $G_{r2}$.
Let $G_s$ and $G_d$ be the antenna array gain of source and destination, respectively.
Furthermore, let $\mathbb{G}_{\textrm{sr}}$ and $\mathbb{G}_{\textrm{dr}}$ be the directivity gain of source-to-relay and relay-to-destination, respectively.
We also define $\theta_s$, $\theta_d$, $\theta_{\textrm{rs}}$, and $\theta_{\textrm{rd}}$ as the instantaneous orientation deviations of source, destination, relay to source, and relay to destination, respectively.

{\bf Theorem 2.}
{\it
For the considered U2U2U link, assuming a fixed gain AF relaying scheme\footnote{{A fixed gain AF relaying system is a relatively simple protocol in which the relay nodes only amplify the source signal with a fixed gain and forward it to the destination without performing any decoding process. For AF relaying, the relay nodes do not require instantaneous channel state information to control the gain of its amplifiers \cite{waqar2010exact}. Due to its simplicity, it is suitable for UAV relaying protocol.}},
the analytical expressions for the PDF of end-to-end SNR at the destination node is given as
\begin{align}
\label{pl1}
&f_{\gamma_{\textrm{sd}}}(\gamma_{\textrm{sd}})
= \int_{\gamma_{\textrm{sd}}}^\infty \sum_{i=0}^{M-1} \sum_{j=0}^{M-1}  \sum_{k=0}^{M-1}
K_3(i,j,k)
\frac{\gamma_{\textrm{dr}}^{2 m}\gamma_{\textrm{sd}}^{m-1}}{\left(\gamma_{\textrm{dr}}-\gamma_{\textrm{sd}} \right)^{m+1}} \nonumber \\
& \times
\exp\left(- \frac{\gamma_{\textrm{dr}}}{N \cos\left(\frac{\pi N j}{2M N} \right)^{2.5} }
\left[K_1(k)
+ \frac{K_1(i) \gamma_{\textrm{sd}} }
{(\gamma_{\textrm{dr}}-\gamma_{\textrm{sd}})} \right]\right) d\gamma_{\textrm{dr}} ,
\end{align}
where
\begin{align}
&K_3(i,j,k) =
\frac{  K_1(k)^{m} K_2(i)
	A_{\textrm{d}k}\left(\theta'_{d},\sigma_{d}\right)
	A_{\textrm{R}j}\left(\theta'_R,\sigma_R\right) }
{\left(N \cos\left(\frac{\pi N j}{2M N} \right)^{2.5} \right)^{2m}\Gamma(m)}, \\
&K_2(i) = \frac{  K_1(i)^{m}  A_{\textrm{s}i}\left(\theta'_{s},\sigma_{s}\right) }{  \Gamma(m)      },  \\
&K_1(i) = \frac{m \sigma^2}{h_L(Z_n) N \cos\left(\frac{\pi N i}{2M N} \right)^{2.5}},
\end{align}
and for $n\in\{\textrm{s,d,R}\}$, the coefficient $A_{ni}$ is obtained from \eqref{a} by replacing $\theta'_{n}$ and $\sigma_{n}$ with $\theta'_{\textrm{ty}}$ and $\sigma_{\textrm{ty}}$, respectively.}
%
%
%
%
%
\\

\begin{IEEEproof}
As shown in Fig. \ref{relay}a, in practice, the orientation deviations of an aerial relay with respect to the source and destination nodes are symmetrical, therefore  $\theta_{\textrm{rs}}=\theta_{\textrm{rd}}=\theta_{R}\sim \mathcal{N}(\theta'_R,\sigma_R^2)$, and consequently $G_{r1}(\theta_R)=G_{r2}(\theta_R)=G_r(\theta_R)$. Thus, the PDF of directivity gain of source-to-relay and relay-to-destination conditioned on
$G_r$ can be obtained as
\begin{align}
\label{sc1}
f_{\mathbb{G}_{nr|G_{r}}}(\mathbb{G}_{nr})
=& \sum_{i=0}^{M-1} \frac{A_{ni}\left(\theta'_{n},\sigma_{n}\right)}{G_{r}}  \\
&\times \delta\left(\frac{\mathbb{G}_{nr}}{G_{r}} - N \cos\left(\frac{\pi N i}{2M N} \right)^{2.5} \right), \nonumber
\end{align}
where the subscript $n\in\{\textrm{s,d}\}$ denotes, respectively, the source-to-relay and relay-to-destination links, and $A_{ni}$ is obtained from \eqref{a} by replacing $\theta'_{n}$ and $\sigma_{n}$ with $\theta'_{\textrm{ty}}$ and $\sigma_{\textrm{ty}}$, respectively.
Also, we have
\begin{align}
\label{po33}
f_{G_r}(G_r) =& \sum_{i=0}^{M-1} A_{\textrm{R}i}\left(\theta'_R,\sigma_R\right)  \\
&\times \delta\left(G_{r} - N \cos\left(\frac{\pi N i}{2M N} \right)^{2.5} \right),\nonumber
\end{align}
where $A_{\textrm{R}i}$ is obtained from \eqref{a} by replacing $\theta'_{r}$ and $\sigma_{r}$ with $\theta'_{\textrm{ty}}$ and $\sigma_{\textrm{ty}}$, respectively.

Assuming a fixed gain AF relaying scheme,
the end-to-end SNR can be obtained from \cite{hasna2002performance}
\begin{align}
\label{sddf}
\gamma_{\textrm{sd}} = \frac{\gamma_{\textrm{sr}} \gamma_{\textrm{dr}}}{\gamma_{\textrm{sr}}+\gamma_{\textrm{dr}}},
\end{align}
where $\gamma_{\textrm{sr}}= \frac{\zeta_s h_L(Z_s) \mathbb{G}_{\textrm{sr}}}{\sigma^2}$ and
$\gamma_{\textrm{dr}}= \frac{\zeta_d h_L(Z_d) \mathbb{G}_{\textrm{dr}}}{\sigma^2}$ represent the SNR at the relay node and destination node, respectively.
Moreover, $\zeta_s$ and $\zeta_d$ denote the small-scale fading and $Z_s$ and $Z_d$ are the link length between source to relay and relay to destination, respectively. Using \eqref{sc1}, for $n\in\{s,d\}$, we can find the PDF of $\gamma_{\textrm{nr}}$ conditioned on $G_r$, as follows:
\begin{align}
\label{xc1}
&f_{\gamma_{\textrm{nr}|G_r}}(\gamma_{nr}) = \int f_{\gamma_{nr}|G_r,\mathbb{G}_{nr}}(\gamma_{nr}) f_{\mathbb{G}_{nr}|G_r}(\mathbb{G}_{nr})d\mathbb{G}_{nr}   \\
&= \int    \frac{\sigma^2}{h_L(Z_n) \mathbb{G}_{nr}}   f_\zeta \left( \frac{\sigma^2\gamma_{nr}}{h_L(Z_n) \mathbb{G}_{nr}} \right)
f_{\mathbb{G}_{nr}|G_r}(\mathbb{G}_{nr})d\mathbb{G}_{nr} \nonumber \\
&= \sum_{i=0}^{M-1} \int
A_{ni}\left(\theta'_{n},\sigma_{n}\right)
\frac{  \left( \frac{m \sigma^2}{h_L(Z_n) \mathbb{G}_{nr}} \right)^{m}}{\Gamma(m)} \nonumber \\
&~~~\times \gamma_{nr}^{m-1}
\exp\left(- \left( \frac{m \sigma^2\gamma_{nr}}{h_L(Z_n) \mathbb{G}_{nr}} \right)\right)   \nonumber \\
&~~~\times \delta\left(\mathbb{G}_{nr} - N G_r\cos\left(\frac{\pi N i}{2M N} \right)^{2.5} \right)  d\mathbb{G}_{nr} \nonumber \\
&= \sum_{i=0}^{M-1}
A_{ni}\left(\theta'_{n},\sigma_{n}\right)
\frac{  K_1(i)^{m}}{G_r^m\Gamma(m)}   \gamma_{nr}^{m-1}
\exp\left(- \frac{K_1(i) \gamma_{nr}}{G_r}\right),  \nonumber
\end{align}
where $K_1(i) = \frac{m \sigma^2}{h_L(Z_n) N \cos\left(\frac{\pi N i}{2M N} \right)^{2.5}}$.
Now, from \eqref{sddf} and \eqref{xc1}, for $\gamma_{\textrm{dr}} > \gamma_{\textrm{sd}}$, the PDF of $\gamma_{\textrm{sd}}$ conditioned on $\gamma_{\textrm{dr}}$ and $G_r$ is obtained as
\begin{align}
\label{zv}
&f_{\gamma_{\textrm{sd}}|\gamma_{\textrm{dr}},G_r}(\gamma_{\textrm{sd}}) = \frac{d}{d\gamma_{\textrm{sd}}}
{\textrm{Pr}}\left\{ \frac{\gamma_{\textrm{sr}} \gamma_{\textrm{dr}}}{\gamma_{\textrm{sr}}+\gamma_{\textrm{dr}}} <\gamma_{\textrm{sd}} \Big |\gamma_{\textrm{dr}},G_r \right\}  \nonumber\\
&= \frac{d}{d\gamma_{\textrm{sd}}}
{\textrm{Pr}}\left\{ 0<\gamma_{\textrm{sr}}< \frac{\gamma_{\textrm{sd}} \gamma_{\textrm{dr}}}{\gamma_{\textrm{dr}}-\gamma_{\textrm{sd}}} \Big |\gamma_{\textrm{dr}},G_r \right\}  \\
& = \frac{\gamma_{\textrm{dr}}^2}{\left( \gamma_{\textrm{dr}}-\gamma_{\textrm{sd}} \right)^2}
f_{\gamma_{\textrm{sr}}|G_r}\left(    \frac{\gamma_{\textrm{sd}} \gamma_{\textrm{dr}}}{\gamma_{\textrm{dr}}-\gamma_{\textrm{sd}}}   \right) \nonumber \\
& = \sum_{i=0}^{M-1}
\frac{K_2(i)}{G_r^m}
\frac{\gamma_{\textrm{dr}}^{m+1}\gamma_{\textrm{sd}}^{m-1}}{\left(\gamma_{\textrm{dr}}-\gamma_{\textrm{sd}} \right)^{m+1}}
\exp\left(-      \frac{K_1(i) \gamma_{\textrm{dr}}\gamma_{\textrm{sd}} }{G_r(\gamma_{\textrm{dr}}-\gamma_{\textrm{sd}})} \right), \nonumber
\end{align}
where
\begin{align}
K_2(i) = \frac{  K_1(i)^{m}  A_{\textrm{s}i}\left(\theta'_{s},\sigma_{s}\right) }{  \Gamma(m)      }.
\nonumber
\end{align}
and $\textrm{Pr}\{\cdot\}$ represents the probability of the event. Finally, from \eqref{po33}, \eqref{xc1}, and \eqref{zv}, the PDF of $\gamma_{sd}$ is obtained in \eqref{pl1}.
%
%
%
%
\end{IEEEproof}
%
%
%
%
%

From \eqref{pl1}, the system performance metrics for a U2U2U link, e.g., channel capacity, and bit error rate can be analytically
characterized without resorting to time-consuming simulations.
Note that the outage probability is obtained by studying the behavior of $F_{\gamma_{\textrm{sd}}}(\gamma_{\textrm{sd}})$ for the lower values of $\gamma_{\textrm{sd}}$.
In the following, we derive a simpler closed-form expression for the proposed channel model that is valid for calculating the outage probability.\\

{\bf Proposition 1.} 
{\it For the considered U2U2U link, the analytical expressions for the CDF of end-to-end SNR at the destination is given by:
\begin{align}
\label{plq1}
&F_{\gamma_{\textrm{sd}}}(\gamma_{\textrm{sd}}) = \sum_{j=0}^{M-1} A_{\textrm{R}j}\left(\theta'_R,\sigma_R\right)  \\
&\times \left[  \sum_{i=0}^{M-1}
\frac{A_{\textrm{s}i}\left(\theta'_{s},\sigma_{s}\right) }{     \Gamma(m)      }
\mathbb{V}\left(m,\frac{K_1(i)}{N \cos\left(\frac{\pi N j}{2M N} \right)^{2.5}} \gamma_{\textrm{sr}}\right) \right. \nonumber \\
&+\sum_{k=0}^{M-1}
\frac{A_{\textrm{d}i}\left(\theta'_{d},\sigma_{d}\right) }{     \Gamma(m)      }
\mathbb{V}\left(m,\frac{K_1(k)}{N \cos\left(\frac{\pi N j}{2M N} \right)^{2.5}} \gamma_{\textrm{dr}}\right) \nonumber \\
&-\sum_{i=0}^{M-1}  \sum_{k=0}^{M-1}
\frac{A_{\textrm{s}i}\left(\theta'_{s},\sigma_{s}\right) }{     \Gamma(m)      }
\mathbb{V}\left(m,\frac{K_1(i)}{N \cos\left(\frac{\pi N j}{2M N} \right)^{2.5}} \gamma_{\textrm{sr}}\right) \nonumber \\
&\left. \times ~
\frac{A_{\textrm{d}i}\left(\theta'_{d},\sigma_{d}\right) }{     \Gamma(m)      }
\mathbb{V}\left(m,\frac{K_1(k)}{N \cos\left(\frac{\pi N j}{2M N} \right)^{2.5}} \gamma_{\textrm{dr}}\right) \right]. \nonumber
\end{align}
}

\begin{figure}
	\begin{center}
		\includegraphics[width=3.35 in]{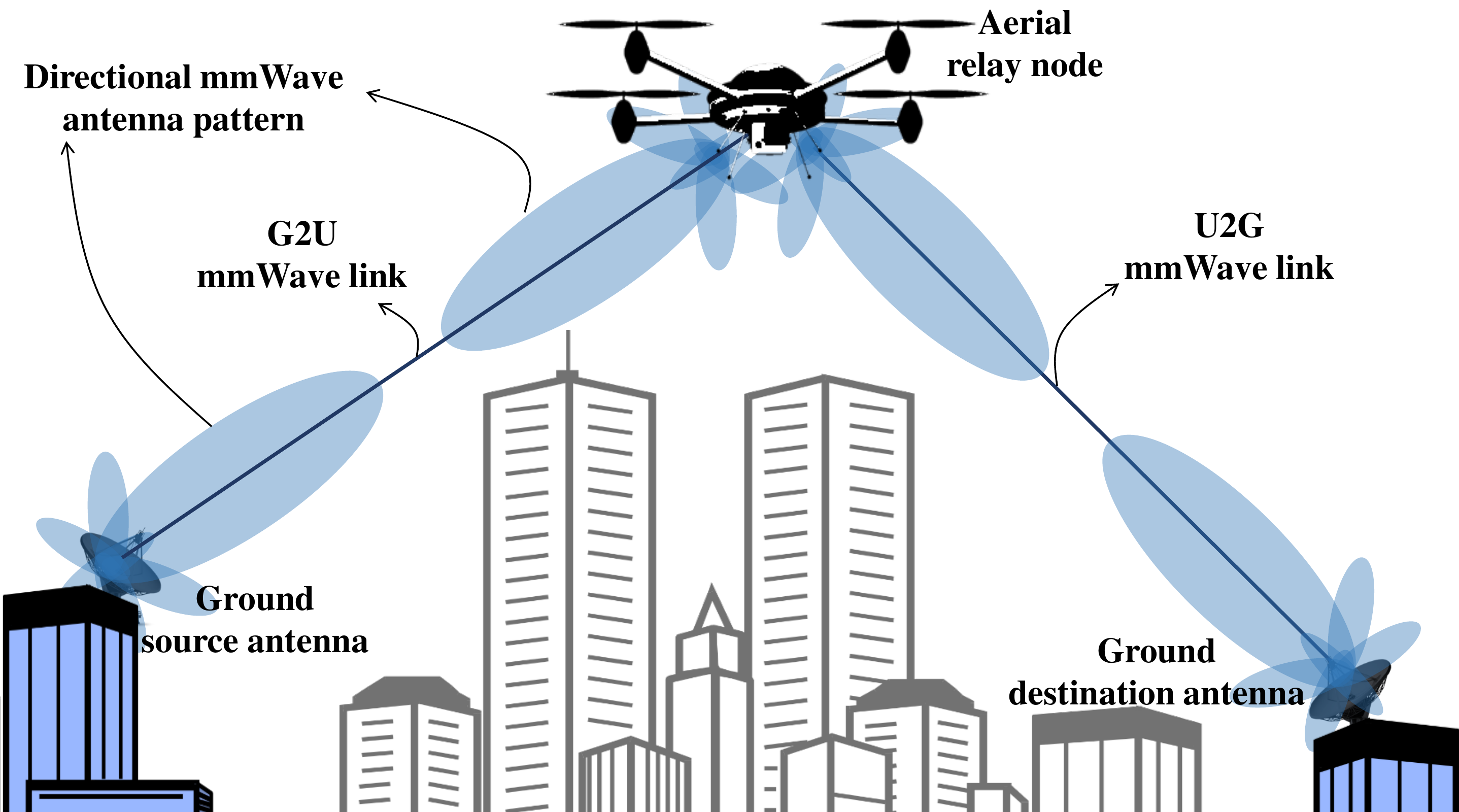}
		\caption{Illustration of a high bandwidth mmWave G2U2G link when the LoS path of source-to-destination link is blocked by tall buildings. G2U and U2G denote the ground source to aerial relay and aerial relay to ground destination links, respectively.}
		\label{gug}
	\end{center}
\end{figure}
%
\begin{IEEEproof}
By using the upper bound approximation given in \cite{ikki2010performance}, we can approximate $\gamma_{\textrm{sd}}$ close to the origin as
\begin{align}
\label{ps1}
\gamma_{\textrm{sd}}\simeq{\rm min}\big(\gamma_{\textrm{sr}},\gamma_{\textrm{dr}}\big).
\end{align}
According to \eqref{ps1}, the CDF of $\gamma_{\textrm{sd}}$ conditioned on $G_r$ can be obtained as
\begin{align}
\label{ps2}
F_{\gamma_{\textrm{sd}}|G_r}(\gamma_{\textrm{sd}}) &= {\textrm{Pr}}\left\{ {\rm min}\big(\gamma_{\textrm{sr}},\gamma_{\textrm{dr}}\big)< \gamma_{\textrm{sd}}\big| G_r \right\}  \\
&={\textrm{Pr}}\!\left\{ \gamma_{\textrm{sr}}\!<\! \gamma_{\textrm{sd}}\big| G_r \!\right\}
\! +{\textrm{Pr}}\!\left\{ \gamma_{\textrm{dr}}\!< \gamma_{\textrm{sd}}\big| G_r \right\} \nonumber \\
&\,\,\,\,\,\,-{\textrm{Pr}}\left\{\gamma_{\textrm{sr}}< \gamma_{\textrm{sd}}\big| G_r \right\}
  {\textrm{Pr}}\left\{\gamma_{\textrm{dr}}< \gamma_{\textrm{sd}}\big| G_r \right\}. \nonumber
\end{align}
From \eqref{xc1} and \eqref{ps2} and using \cite[(3.381.1)]{jeffrey2007table}, we have
\begin{align}
\label{pa1}
F_{\gamma_{\textrm{sd}}|G_r}(\gamma_{\textrm{sd}}) =& ~F_{\gamma_{\textrm{sr}}|G_r}(\gamma_{\textrm{sr}}) + F_{\gamma_{\textrm{dr}}|G_r}(\gamma_{\textrm{dr}})  \\
&\,\,\,- F_{\gamma_{\textrm{sr}}|G_r}(\gamma_{\textrm{sr}})
F_{\gamma_{\textrm{dr}}|G_r}(\gamma_{\textrm{dr}}),\nonumber
\end{align}
where for $n\in\{s,d\}$,
\begin{align}
\label{gf1}
F_{\gamma_{nr}|G_r}(\gamma_{nr}) =
\sum_{i=0}^{M-1}
\frac{A_{ni}\left(\theta'_{n},\sigma_{n}\right) }{     \Gamma(m)      }
\mathbb{V}\left(m,\frac{K_1(i)}{G_r} \gamma_{nr}\right).
\end{align}
Finally, from \eqref{po33}, \eqref{pa1} and \eqref{gf1}, the CDF of $\gamma_{\textrm{sd}}$ is derived in \eqref{plq1}.
\end{IEEEproof}
From the simpler expression for the CDF of the end-to-end SNR of a U2U2U link provided in \eqref{plq1}, the outage probability for such link can be analytically developed without simulations.
%
%
%
%
\subsection{G2U2G Link}
An shown in Fig. \ref{gug}, G2U2G link is a special case of an aerial relay-assisted communication system for which source and destination are ground nodes. Next, the PDF and the CDF of the instantaneous SNR of the G2U2G link are derived. \\

{\bf Proposition 2.}
{\it
For the considered G2U2G link, the analytical expressions for the PDF and the CDF of end-to-end SNR at the destination are respectively given by:
\begin{align}
\label{xw2}
f_{\gamma_{\textrm{sd}}}(\gamma_{sd}) =
&\sum_{i=0}^{M-1} A_{\textrm{R}i}\left(\theta'_R,\sigma_R\right)\frac{\sqrt{\pi} B'' }{2^{2m-1}\Gamma^2(m)         }  \\
&\times G_{1,2}^{2,0}\left(B''  \gamma_{\textrm{sd}}   \Bigg|  {m-\frac{1}{2} \atop  m-1, 2m-1} \right),  ~~~\gamma_{\textrm{sd}}>0  \nonumber
\end{align}	
and
\begin{align}
\label{cdf3}
F_{\gamma_{\textrm{sd}}}(\gamma_{\textrm{sd}}) =&
\sum_{i=0}^{M-1} A_{\textrm{R}i}\left(\theta'_R,\sigma_R\right)\frac{\sqrt{\pi} B'' \gamma_{\textrm{sd}}  }{2^{2m-1}\Gamma^2(m)         }  \\
&\times G_{2,3}^{2,1}\left(B''  \gamma_{\textrm{sd}}   \Bigg|  {0, m-\frac{1}{2} \atop  m-1, 2m-1, -1} \right). \nonumber
\end{align}
where $B'' =  \frac{4 m \sigma^2}{N^2 h_L(Z) \cos\left(\frac{\pi N i}{2M N} \right)^{2.5} }$
and $G_{m,n}^{p,q}$ is the Meijer's G-function \cite{jeffrey2007table}.}

\begin{IEEEproof}
In the G2U2G link, as shown in Fig. \ref{relay2}, it is reasonable to assume that the ground nodes are firmly fixed, and, hence, their slight vibrations can be ignored. Given this assumption and from \eqref{gap}, we have $G_s = G_d = N $.
From \eqref{snr1} and similar to the derivations of \cite[Appendix A and D]{hasna2004harmonic}, when $Z_1=Z_2=Z$, the PDF of $\gamma_{\textrm{sd}}$ conditioned on $G_r$ is obtained as
\begin{align}
\label{xw1}
&f_{\gamma_{\textrm{sd}}|G_r}(\gamma_{\textrm{sd}}) =
\frac{\sqrt{\pi}\,m\, \sigma^2}{2^{2m-3}\Gamma^2(m)       N h_L(Z) G_{r}  }  \\
~~~~~~~~&\times G_{1,2}^{2,0}\left( \frac{4 m\gamma_{\textrm{sd}} \sigma^2 }{N h_L(Z) G_{r}} \Bigg|  {m-\frac{1}{2} \atop  m-1, 2m-1} \right),
~~~0\leq \gamma_{\textrm{sd}}.   \nonumber
\end{align}
Finally, we have
\begin{align}
\label{xwoo2}
f_{\gamma_{\textrm{sd}}}(\gamma_{sd}) =& \int   f_{\gamma_{\textrm{sd}}|G_r}(\gamma_{\textrm{sd}})  f_{G_r}(G_r)  dG_r   \\
=&\sum_{i=0}^{M-1} A_{\textrm{R}i}\left(\theta'_R,\sigma_R\right)\frac{\sqrt{\pi} B'' }{2^{2m-1}\Gamma^2(m)         } \nonumber \\
&\times G_{1,2}^{2,0}\left(B''  \gamma_{\textrm{sd}}   \Bigg|  {m-\frac{1}{2} \atop  m-1, 2m-1} \right),  ~~~\gamma_{\textrm{sd}}>0,  \nonumber
\end{align}
where $B'' =  \frac{4 m \sigma^2}{N^2 h_L(Z) \cos\left(\frac{\pi N i}{2M N} \right)^{2.5} }$. Moreover, employing \cite[(7.811.2)]{jeffrey2007table}, the CDF of $\gamma_{\textrm{sd}}$ is derived in \eqref{cdf3}.
\end{IEEEproof}
\begin{figure*}
	\centering
	\subfloat[] {\includegraphics[width=2.3 in]{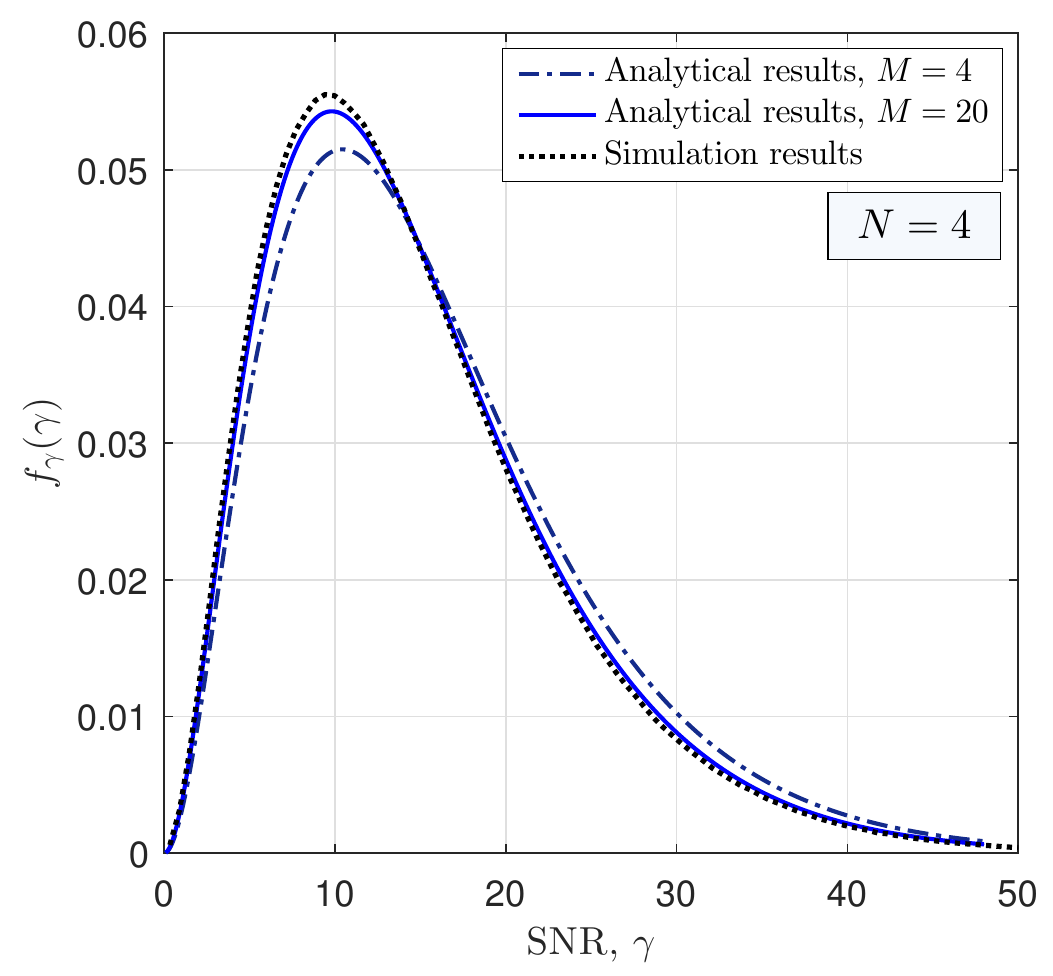}
		\label{x1}
	}
	\hfill
	\subfloat[] {\includegraphics[width=2.3 in]{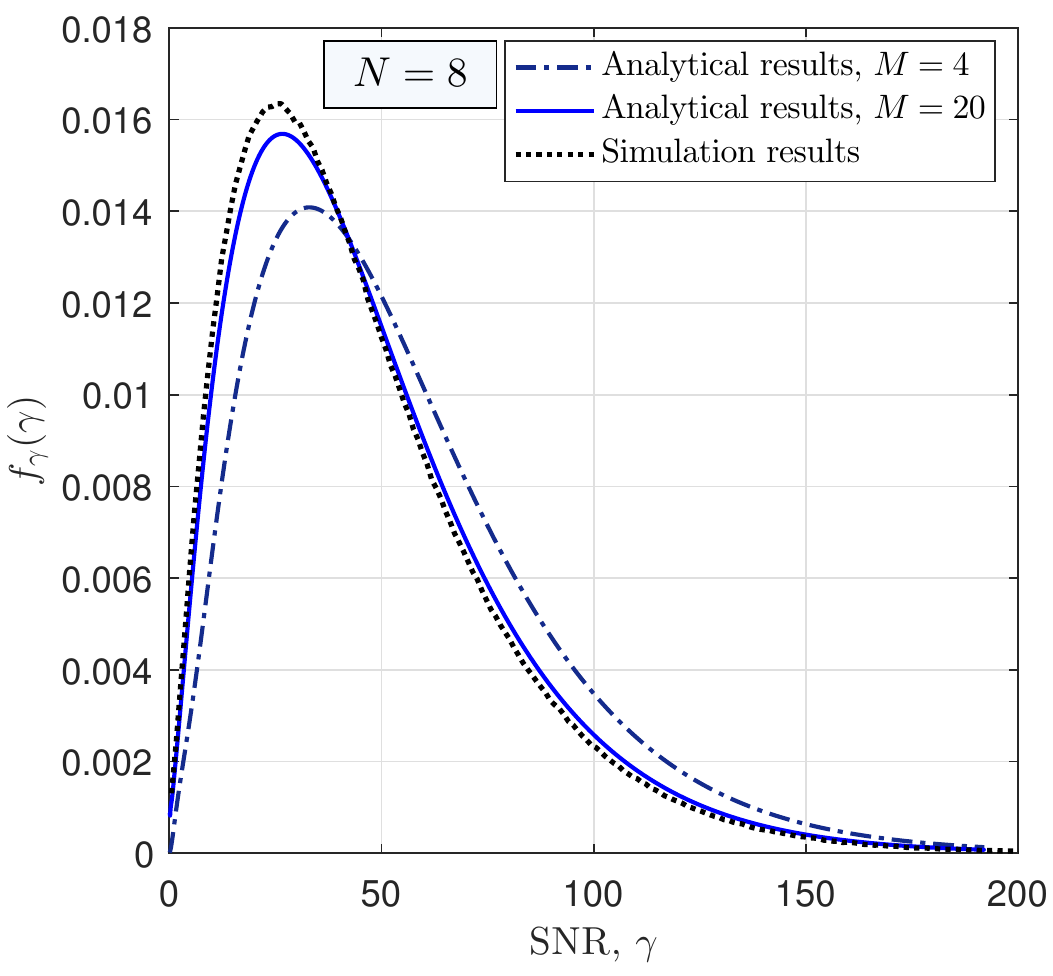}
		\label{x2}
	}
	\hfill
	\subfloat[] {\includegraphics[width=2.3 in]{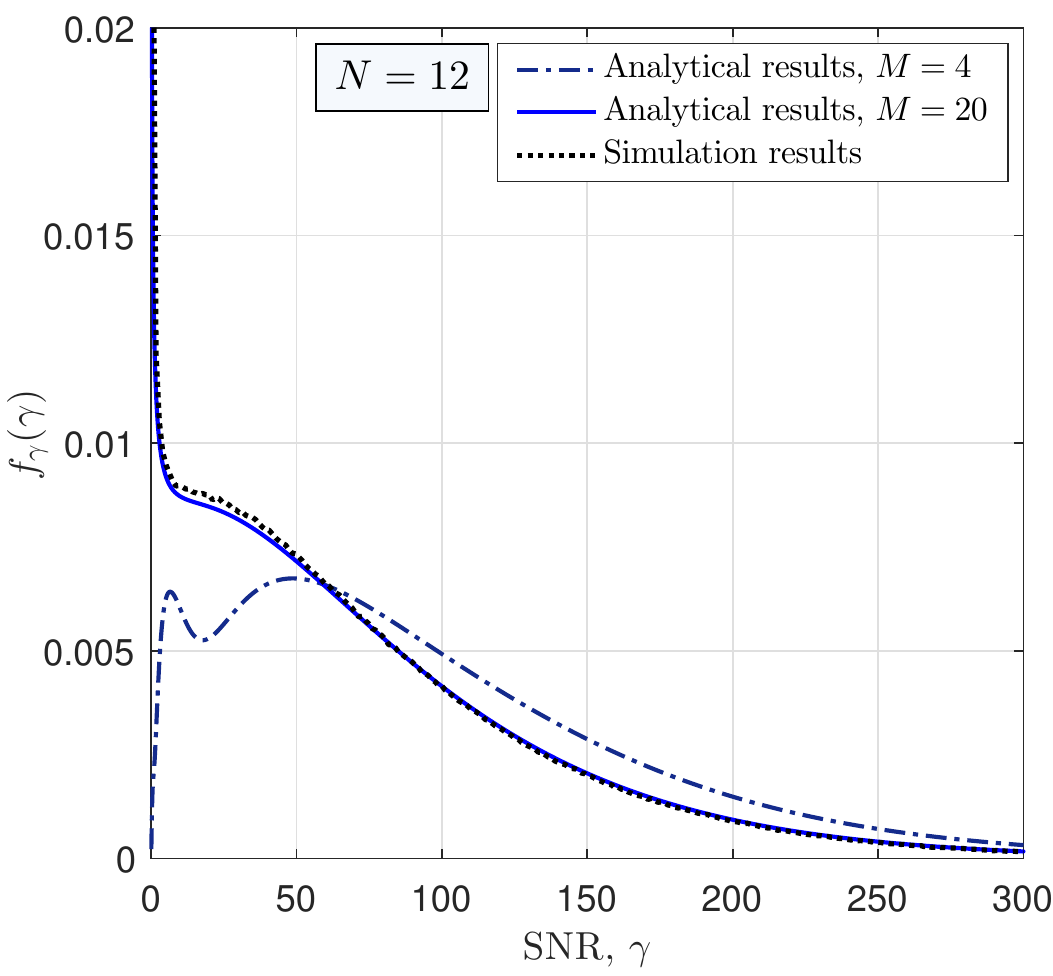}
		\label{x3}
	}
	\caption{Channel distribution of U2U link when $\sigma_{\textrm{ty}}=\sigma_{\textrm{ry}}=30$ $\rm mrad$ and $\theta'_{\textrm{ty}}=\theta'_{\textrm{ry}}=5$ $\rm mrad$ for a) $N=4$, b) $N=8$, and c) $N=12$. }
	\label{x4}
\end{figure*}
%
%
\begin{figure*}
	\centering
	\subfloat[] {\includegraphics[width=2.3 in]{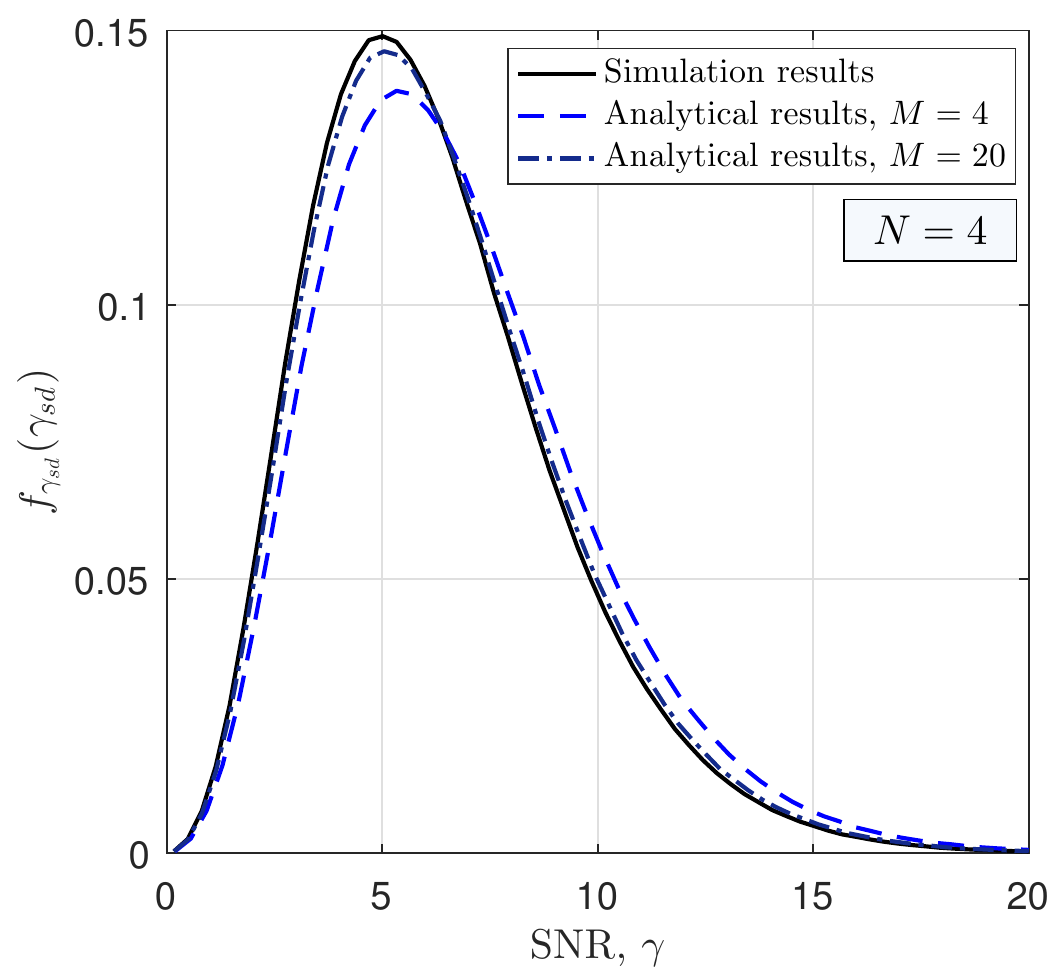}
		\label{xx1}
	}
	\hfill
	\subfloat[] {\includegraphics[width=2.3 in]{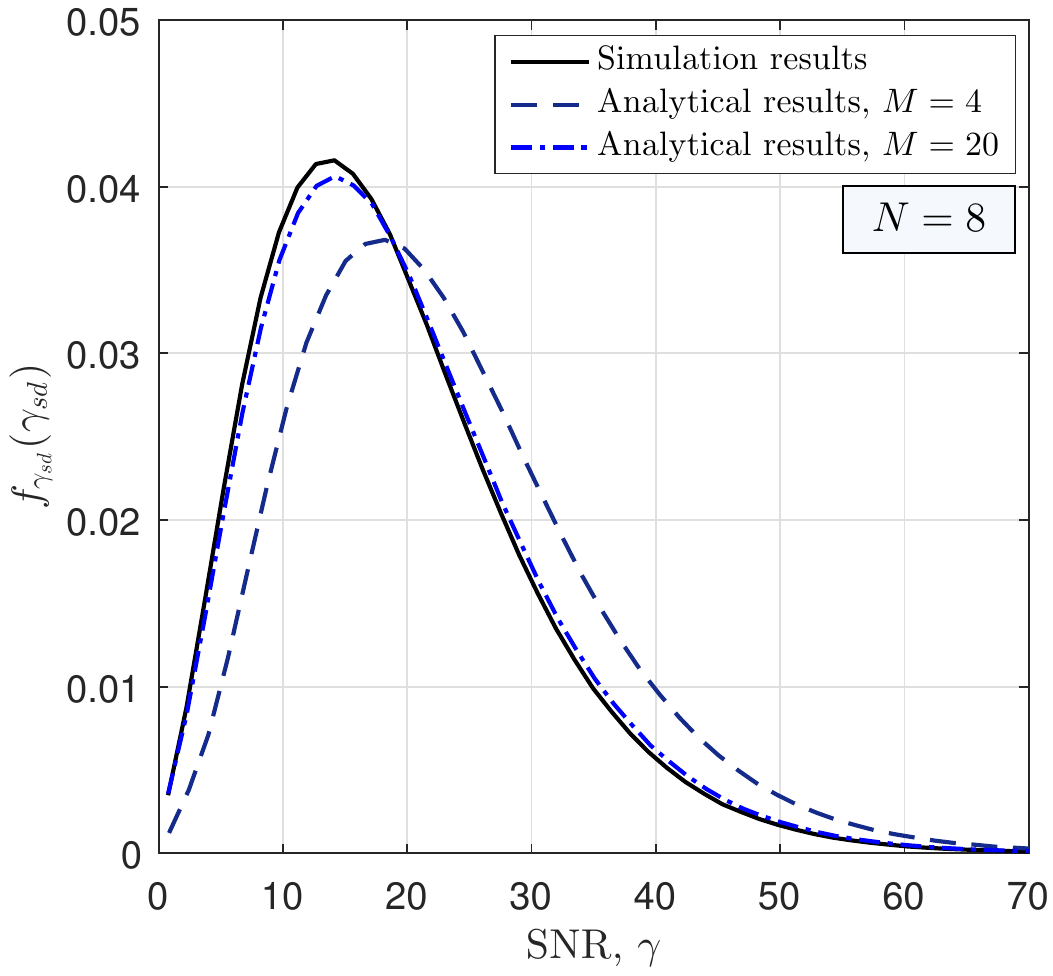}
		\label{xx2}
	}
	\hfill
	\subfloat[] {\includegraphics[width=2.3 in]{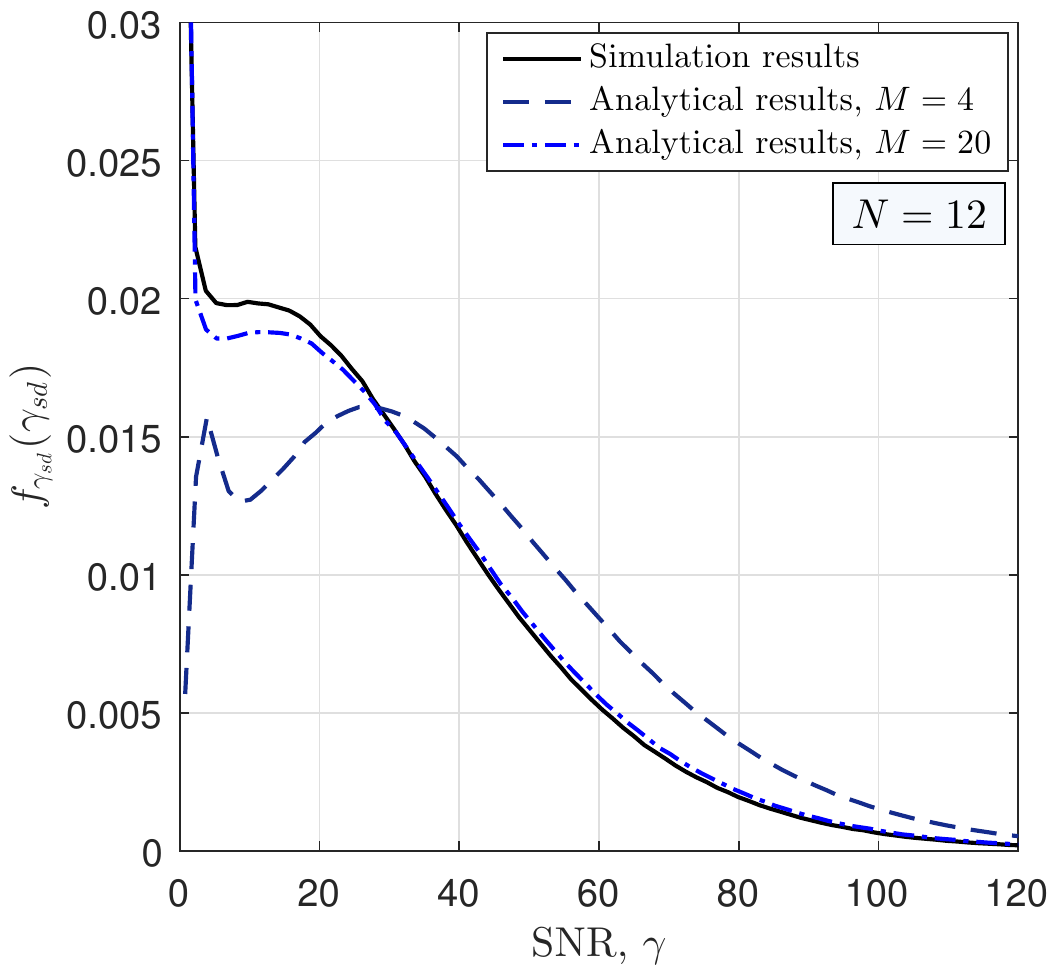}
		\label{xx3}
	}
	\caption{Channel distribution of aerial relay link when $\sigma_{s}=\sigma_{r}=\sigma_{d}=30$ $\rm mrad$ and $\theta'_{s}=\theta'_{r}=\theta'_{d}=5$ $\rm mrad$ for a) $N=4$, b) $N=8$, and c) $N=12$.}
	\label{xx4}
\end{figure*}
%

%
\begin{figure}
	\begin{center}
		\includegraphics[width=3.3 in]{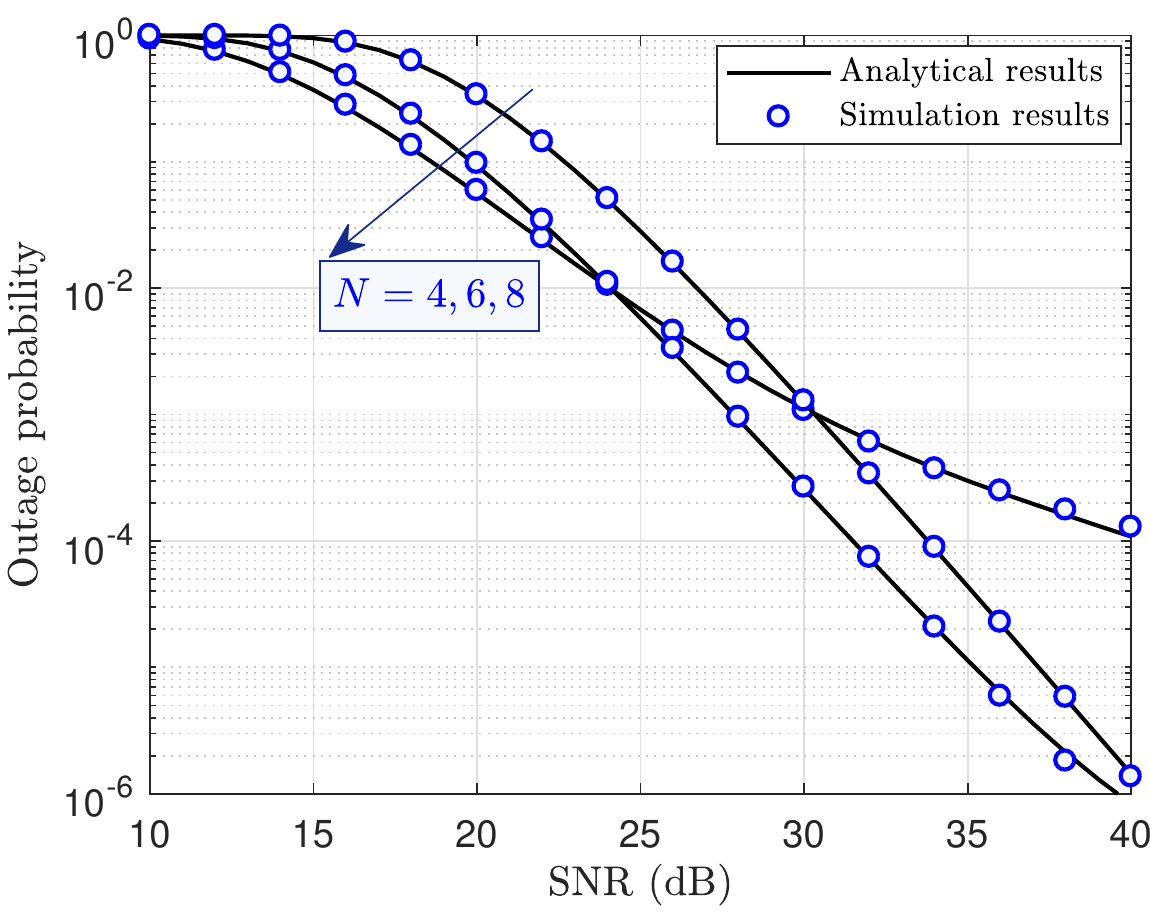}
		\caption{Outage probability of U2U link for  $\sigma_{\textrm{ty}}=\sigma_{\textrm{ry}}=30$ $\rm mrad$, $\theta'_{\textrm{ty}}=\theta'_{\textrm{ry}}=0$ and different values of antenna elements number $N=4, 6$ and $8$. Analytical results are obtained for $M=20$.}
		\label{O1}
	\end{center}
\end{figure}
%
%
\begin{figure}
	\begin{center}
		\includegraphics[width=3.3 in]{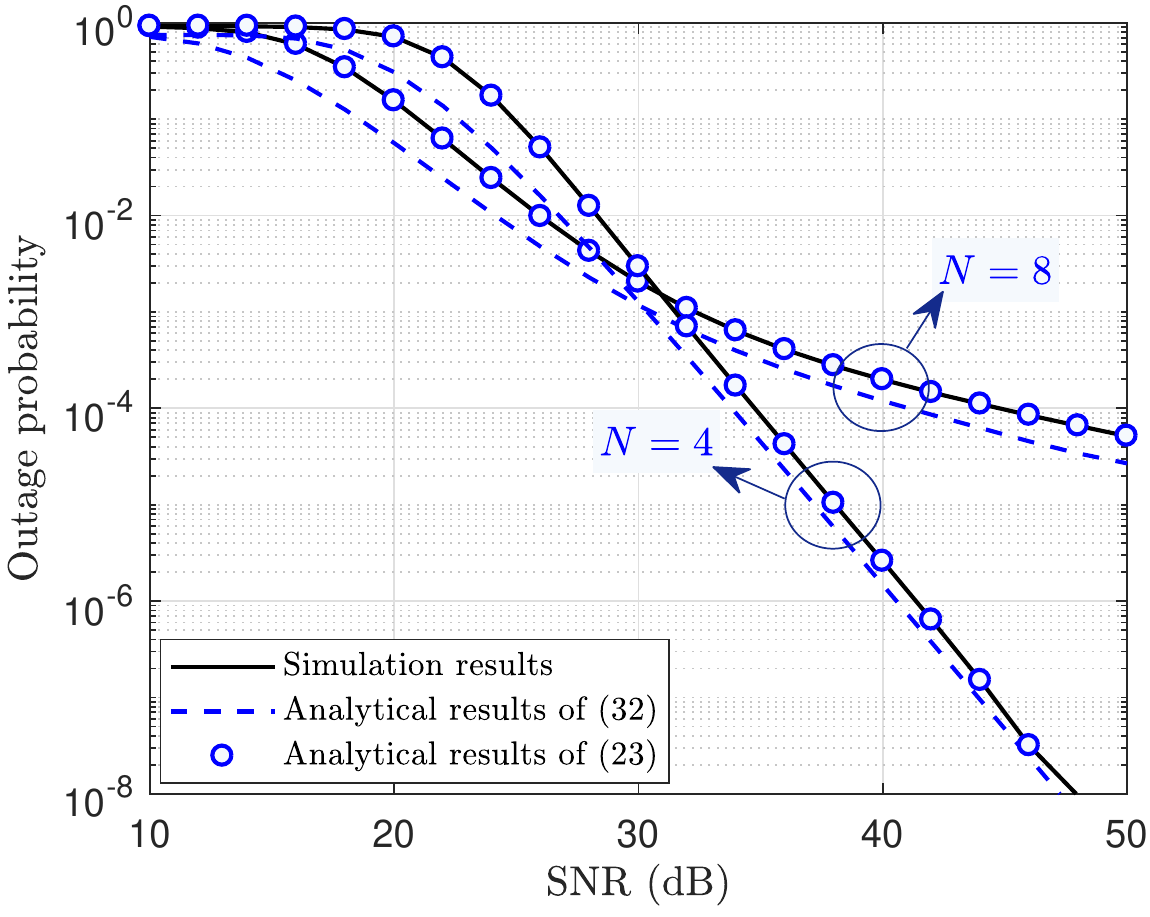}
		\caption{Outage probability of U2U2U link for  $\sigma_{s}=\sigma_{r}=\sigma_{d}=30$ $\rm mrad$ and $\theta'_{s}=\theta'_{r}=\theta'_{d}=0$ and two different values of antenna elements number $N=4$ and $8$. Analytical results are obtained for $M=20$.}
		\label{O2}
	\end{center}
\end{figure}
%
%
\begin{figure}
	\begin{center}
		\includegraphics[width=3.3 in]{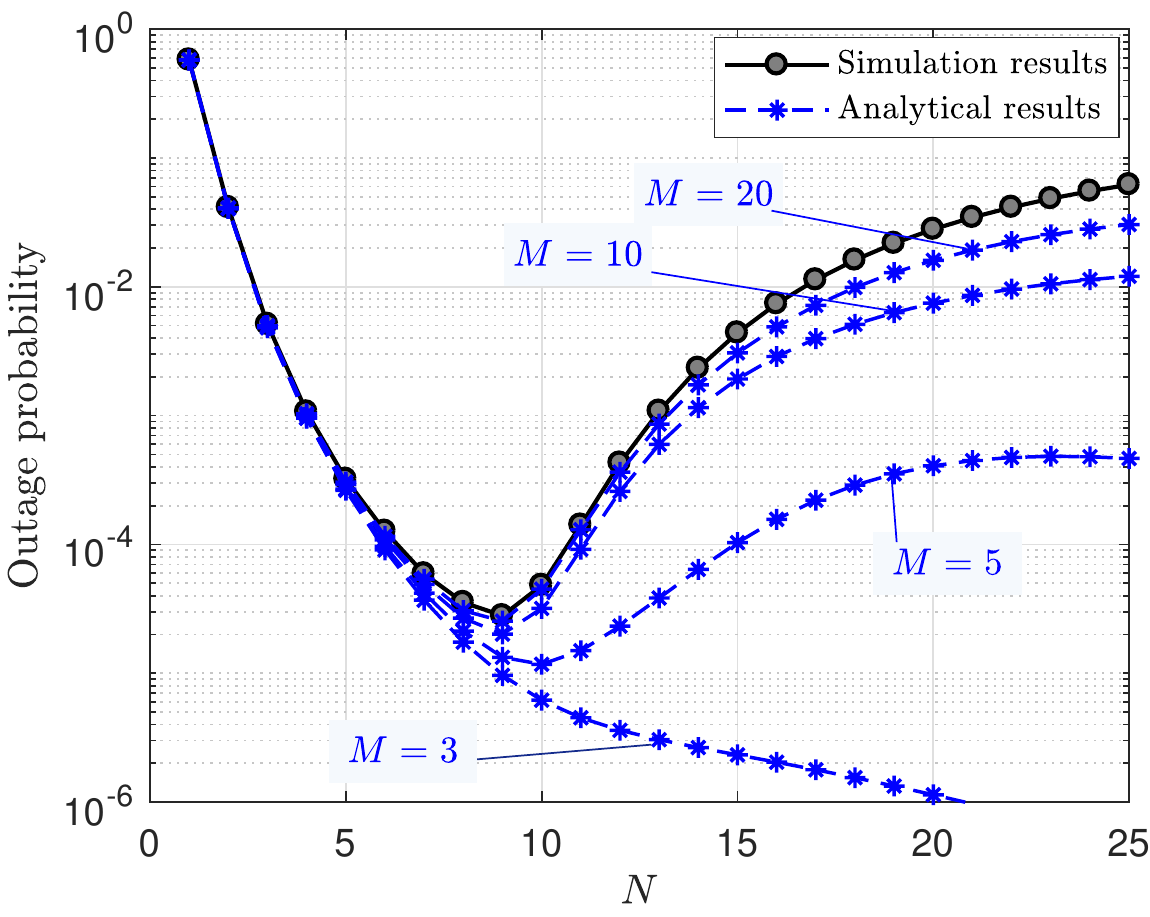}
		\caption{Outage probability of U2U link versus different number of antenna elements $N$ for  $\sigma_{\textrm{ty}}=\sigma_{\textrm{ry}}=20$ $\rm mrad$, $\theta'_{\textrm{ty}}=\theta'_{\textrm{ry}}=0$. Analytical results are obtained for different values of $M$.}
		\label{O3}
	\end{center}
\end{figure}
%
%
\begin{figure}
	\begin{center}
		\includegraphics[width=3.3 in]{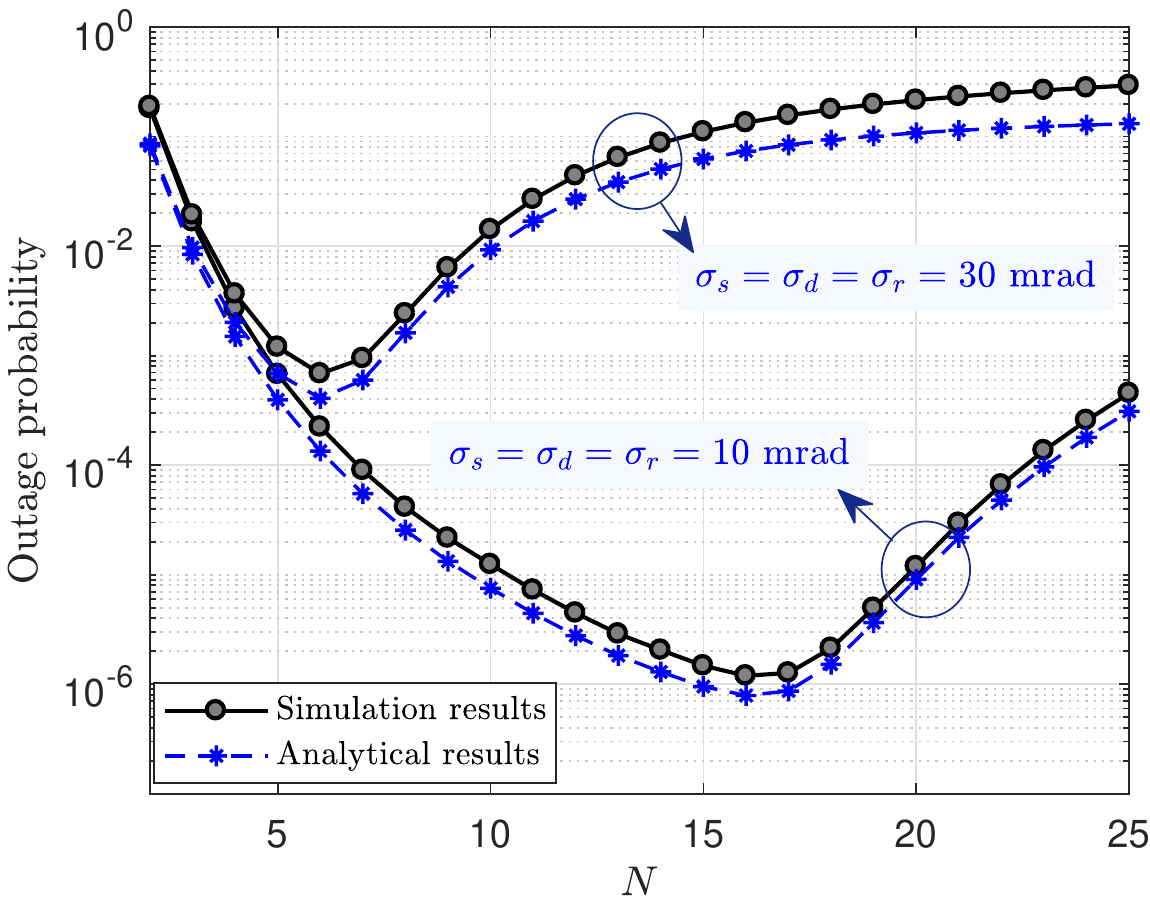}
		\caption{Outage probability of U2U2U link versus different number of antenna elements $N$ for SNR=30 dB,  $\theta'_{s}=\theta'_{r}=\theta'_{d}=0$ and two different insatiability conditions $\sigma_{s}=\sigma_{r}=\sigma_{d}=10$ and $30$ $\rm mrad$. }
		\label{O4}
	\end{center}
\end{figure}
%
%
\begin{figure}
	\begin{center}
		\includegraphics[width=3.3 in]{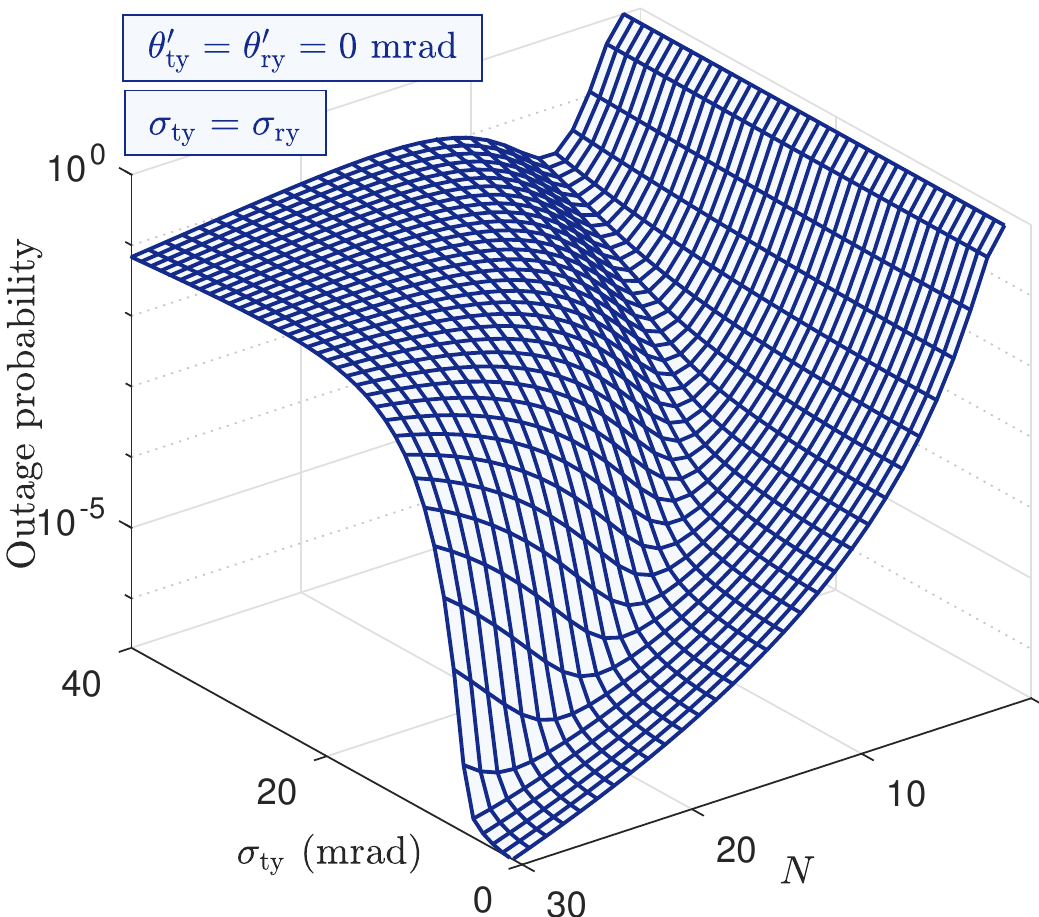}
		\caption{Outage probability of U2U link versus $N$ and $\sigma_{\textrm{ty}}$ for SNR=25 dB when  $\sigma_{\textrm{ty}}=\sigma_{\textrm{ry}}$ and $\theta'_{\textrm{ty}}=\theta'_{\textrm{ry}}=0$. }
		\label{3D_uu}
	\end{center}
\end{figure}
%
%
\begin{figure}
	\begin{center}
		\includegraphics[width=3.3 in]{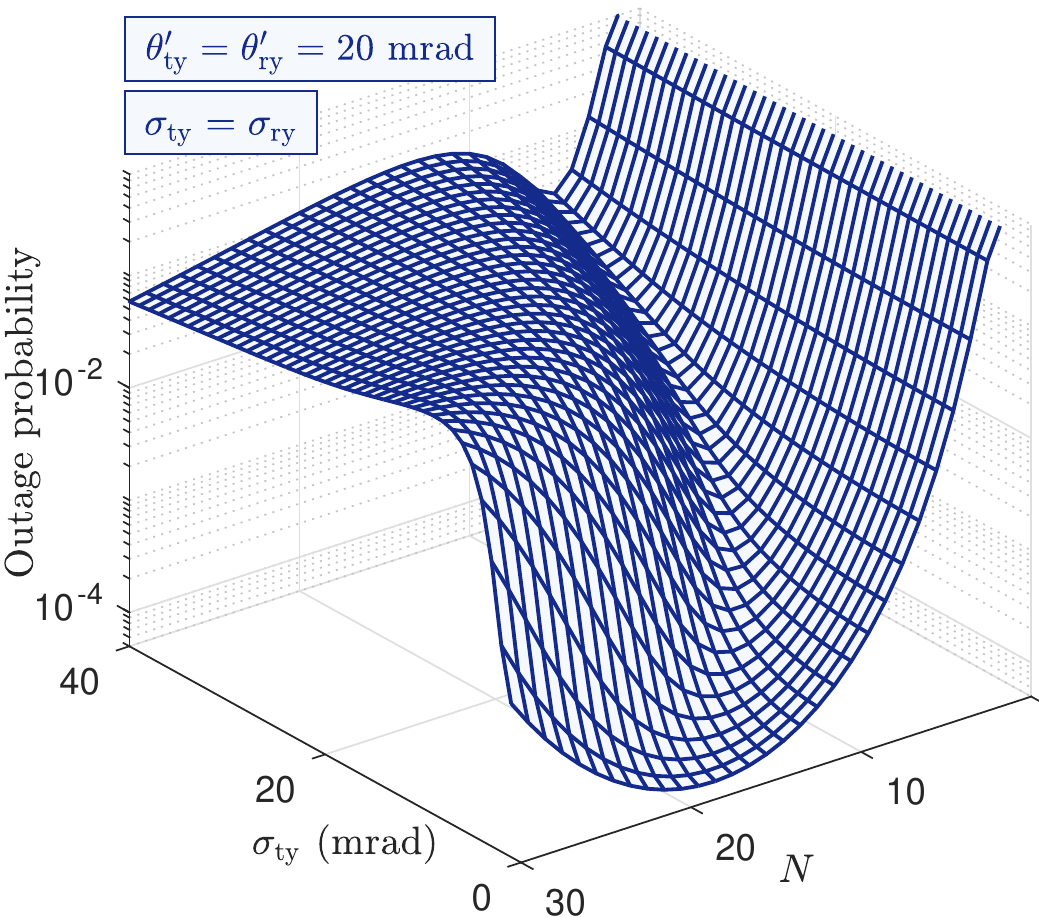}
		\caption{Outage probability of U2U link versus $N$ and $\sigma_{\textrm{ty}}$ for SNR=25 dB when  $\sigma_{\textrm{ty}}=\sigma_{\textrm{ry}}$ and $\theta'_{\textrm{ty}}=\theta'_{\textrm{ry}}=20$ $\rm mrad$.}
		\label{3D_uu2}
	\end{center}
\end{figure}

From \eqref{xw2} and \eqref{cdf3}, the system performance metrics for a G2U2G link, e.g., channel capacity, outage probability, and bit error rate can be tractably characterized.
\section{Simulation Results and Analysis}
For evaluation, we perform Monte Carlo simulations with over $50\times10^6$ independent runs. We corroborate the accuracy of the derived channel model expressions for different values of the parameters related to orientation deviations.
We also evaluate the performance in terms of the outage probability.

For our simulations, we consider the UAVs to have the same standard deviation of AoA and AoD fluctuations, i.e., for the U2U link consist of two hovering UAVs, we have $\sigma_{\textrm{ty}}=\sigma_{\textrm{ry}}$ and for the U2U2U link, we have $\sigma_s=\sigma_d=\sigma_r$.
In the case of the U2U2U link, the link length between the source node to the relay node is assumed to be equal to the link length between the relay node to the destination node. We consider standard values for other system parameters, as follows.
The link length $Z=500$~m, carrier frequency $f_c=60$~GHz, average building height $h_b=25$~m, normalized thermal noise power $\sigma^2=30$~dBm, Nakagami fading parameter $m=3$, and SNR threshold $\gamma_{\textrm{th}}=10$~dB.
\subsection{Accuracy of the Derived Channel Models}
In Figs. \ref{x4} and \ref{xx4}, we show, respectively, the channel distribution of the U2U link and the U2U2U link under the orientations deviations for aerial nodes. To assess the effect of the antenna pattern on the distribution of SNR at the receiver side, results are provided for the different numbers of antenna elements $N$. From these two figures, we can observe that, by increasing the number of antenna elements, the values of the SNR at the receiver will vary within a wider range. Clearly, by increasing the number of antenna elements, the array gain at the center of the beam (main-lobe) becomes narrower and stronger which results in higher SNR values at the cost of higher sensitivity to the beam deviations. To facilitate analysis of such systems, for both links,  the SNR distributions  as a function of $N$ are captured in \eqref{po1}, and \eqref{pl1}, respectively. Moreover, Figs. \ref{x4} and \ref{xx4} show that the accuracy of the analytical results directly depends on the number of sectors $M$, and for sufficiently large values of $M$, an exact match between simulations and theory can be achieved. However, as we can observe from Figs. \ref{x1} and \ref{xx1}, when $N$ is small, the analytical model obtained from $M = 4$ is also reasonable.
Also, the results of Figs. \ref{x3} and \ref{xx3} indicate that, by using small values for $M$, the analytical analysis do not accurately capture the low values of the SNR.
\subsection{Performance Analysis and Optimal Pattern Selection}
Next, we investigate the performance of the considered systems in terms of outage probability.
To demonstrate the impact of changing antenna gain on the system performance at different SNR regimes, in Fig \ref{O1}, the outage probability of U2U link versus SNR for different number of antenna elements, $N$, is presented.
From this figure, we can observe that higher values for $N$ achieve better performance at low SNR regime. Meanwhile, in the high SNR regime, lower values of $N$ result in a more reliable communication link.
On the other hand, at high SNR values, the poor performance of the transceivers with higher directional gain (or more antenna elements) indicates that those transceivers are more vulnerable to the orientation fluctuations due to UAV vibrations. Moreover, the accuracy of the derived closed-form expression for outage probability is verified in Fig. \ref{O1}.

In Fig. \ref{O2}, we study the performance of the U2U2U link by presenting the outage probability as the values of the SNR and the number of antenna elements vary.
{The analytical results provided in Fig. \ref{O2} are obtained based on two different approaches.
First, they are derived by substituting \eqref{pl1} in the integral of \eqref{xd}, and they perfectly match with the simulation results. Second, we provide the analytical results with acceptable accuracy by using \eqref{cdf3} which has a simpler form than the first approach.}

{To shed more light on the importance of antenna pattern optimization, in Figs. \ref{O3} and \ref{O4}, respectively, the outage probability of the U2U link and the outage probability of the U2U2U link versus $N$ are shown.
Fig. \ref{O3} demonstrates that increasing antenna directivity gain (by increasing the number of antenna elements, $N$) does not necessarily improve the system performance. This is expected, since increasing the antenna directivity gain results in a narrower main lobe which makes the link performance more prone to instantaneous vibrations of UAVs.}

Furthermore, from Fig. \ref{O3}, we can observe that the accuracy of the analytical results obtained from \eqref{p1} depends on the values of $M$ and its higher values lead to better accuracy at the expense of increasing computational load.
However, for the optimal value of $N$ that achieves a minimum outage probability, the analytical results of \eqref{p1} for $M=10$ are valid. 
Also, from Fig. \ref{O4}, we can observe that the optimal value of $N$ greatly depends on the instantaneous orientation deviations of UAVs. For instance, the change in  $\sigma_{s}$, $\sigma_{r}$, and $\sigma_{d}$ from 10 $\rm mrad$ to 30 $\rm mrad$, will reduce the optimal value of $N$ from 16 to 6.

%
%
\begin{table}
	\def\tablename{Table}
	\centering
	\caption{Comparison of the Optimal values for $N$ obtained by simulation and numerical results to achieve minimum outage probability over U2U link for different values of $\sigma_{\textrm{ty}}=10$, $20$ and $30$ $\rm mrad$ and two different values of SNR=20  and 30 $\rm dB$ when  $\sigma_{\textrm{ty}}=\sigma_{\textrm{ry}}$ and $\theta'_{\textrm{ty}}=\theta'_{\textrm{ry}}=0$.}
	\begin{tabular}{|c||   c|c|c|c|}
		\cline{2-5}
		\multicolumn{1}{l||}{}&\multicolumn{4}{|c|}{ For SNR=20 dB }\\
		\hline
		\multicolumn{1}{|l||}{	$\sigma_{\textrm{ty}}$} &\multicolumn{2}{|c|}{Simulation results}
		&\multicolumn{2}{|c|}{Analytical results}\\
		\cline{2-5}
		(mrad)
		&Optimal $N$&$\mathbb{P}_{\textrm{out}}$&Optimal $N$&$\mathbb{P}_{\textrm{out}}$\\
		\hline\hline
		10
		&18&$4\times 10^{-4}$         &18&$4\times 10^{-4}$\\
		\hline
		20
		&11&$1.4\times 10^{-2}$         &11&$1.3\times 10^{-2}$\\
		\hline
		30
		&8&$6.5\times 10^{-2}$          &8&$6.3\times 10^{-2}$\\
		\hline
		\hline
		\cline{2-5}
		\multicolumn{1}{l||}{}&\multicolumn{4}{|c|}{For SNR=30 dB}\\
		\hline
		\multicolumn{1}{|l||}{	$\sigma_{\textrm{ty}}$} &\multicolumn{2}{|c|}{Simulation results}
		&\multicolumn{2}{|c|}{Analytical results}\\
		\cline{2-5}
		(mrad)
		&Optimal $N$&$\mathbb{P}_{\textrm{out}}$&Optimal $N$&$\mathbb{P}_{\textrm{out}}$\\
		\hline\hline
		10
		&16&$6.3\times 10^{-7}$         &16&$6.3\times 10^{-7}$\\
		\hline
		20
		&9&$3.4\times 10^{-5}$         &9&$3.4\times 10^{-5}$\\
		\hline
		30
		&6&$3.1\times 10^{-4}$          &6&$3\times 10^{-4}$\\
		\hline
	\end{tabular}
	\label{tab2}%
\end{table}%
%
%
%
\begin{table}
	\def\tablename{Table}
	\centering
	\caption{Comparison of the Optimal values for $N$ obtained by simulation and numerical results to achieve minimum outage probability over U2U2U link for different values of $\sigma_{\textrm{ty}}=10$, $20$ and $30$ $\rm mrad$ and two different values of SNR=20  and 30 $\rm dB$ when  $\sigma_{s}=\sigma_{r}=\sigma_{d}$ and $\theta'_{s}=\theta'_{r}=\theta'_{d}=0$.}
	\begin{tabular}{|c||   c|c|c|c|}
		\cline{2-5}
		\multicolumn{1}{l||}{}&\multicolumn{4}{|c|}{ For SNR=20 dB }\\
		\hline
		\multicolumn{1}{|l||}{	$\sigma_{s}$} &\multicolumn{2}{|c|}{Simulation results}
		&\multicolumn{2}{|c|}{Analytical results}\\
		\cline{2-5}
		(mrad)
		&Optimal $N$&$\mathbb{P}_{\textrm{out}}$&Optimal $N$&$\mathbb{P}_{\textrm{out}}$\\
		\hline\hline
		10
		&18&$5\times 10^{-4}$        &18&$3.8\times 10^{-4}$\\
		\hline
		20
		&11&$3.1\times 10^{-2}$         &11&$1.2\times 10^{-2}$\\
		\hline
		30
		&8&$7.2\times 10^{-2}$          &8&$5.9\times 10^{-2}$\\
		\hline
		\hline
		\cline{2-5}
		\multicolumn{1}{l||}{}&\multicolumn{4}{|c|}{For SNR=30 dB}\\
		\hline
		\multicolumn{1}{|l||}{	$\sigma_{s}$} &\multicolumn{2}{|c|}{Simulation results}
		&\multicolumn{2}{|c|}{Analytical results}\\
		\cline{2-5}
		(mrad)
		&Optimal $N$&$\mathbb{P}_{\textrm{out}}$&Optimal $N$&$\mathbb{P}_{\textrm{out}}$\\
		\hline\hline
		10
		&16&$6.4\times 10^{-7}$        &16&$6.1\times 10^{-7}$\\
		\hline
		20
		&9&$3.9\times 10^{-5}$         &9&$3.2\times 10^{-5}$\\
		\hline
		30
		&6&$3.8\times 10^{-4}$          &6&$2.9\times 10^{-4}$\\
		\hline
	\end{tabular}
	\label{tab3}%
\end{table}%
%
%
\begin{table}
	\def\tablename{Table}
	\centering
	\caption{ Comparison of the Optimal values for $N$ obtained by simulation and numerical results to achieve minimum outage probability over U2U link for different values of $\theta'_{\textrm{ty}}$  and two different values of SNR=20  and 30 $\rm dB$ when  $\sigma_{\textrm{ty}}=\sigma_{\textrm{ry}}=10$ $\rm mrad$ and $\theta'_{\textrm{ty}}=\theta'_{\textrm{ry}}$.}
	\begin{tabular}{|c||   c|c|c|c|}
		\cline{2-5}
		\multicolumn{1}{l||}{}&\multicolumn{4}{|c|}{ For SNR=20 dB }\\
		\hline
		\multicolumn{1}{|l||}{	$\theta'_{\textrm{ty}}$} &\multicolumn{2}{|c|}{Simulation results}
		&\multicolumn{2}{|c|}{Analytical results}\\
		\cline{2-5}
		(mrad)
		&Optimal $N$&$\mathbb{P}_{\textrm{out}}$&Optimal $N$&$\mathbb{P}_{\textrm{out}}$\\
		\hline\hline
		5
		&17&$6.5\times 10^{-4}$         &17&$6.5\times 10^{-4}$ \\
		\hline
		10
		&15&$1.6\times 10^{-3}$         &15&$1.6\times 10^{-3}$\\
		\hline
		15
		&13&$3.9\times 10^{-3}$          &13&$3.8\times 10^{-3}$\\
		\hline
		20
		&12&$8.6\times 10^{-3}$          &12&$8.4\times 10^{-3}$\\
		\hline
		\hline
		\cline{2-5}
		\multicolumn{1}{l||}{}&\multicolumn{4}{|c|}{For SNR=30 dB}\\
		\hline
		\multicolumn{1}{|l||}{	$\theta'_{\textrm{ty}}$} &\multicolumn{2}{|c|}{Simulation results}
		&\multicolumn{2}{|c|}{Analytical results}\\
		\cline{2-5}
		(mrad)
		&Optimal $N$&$\mathbb{P}_{\textrm{out}}$&Optimal $N$&$\mathbb{P}_{\textrm{out}}$\\
		\hline\hline
		5
		&15&$10^{-6}$                   &15&$ 10^{-6}$\\
		\hline
		10
		&14&$2.3\times 10^{-6}$         &14&$2.3\times 10^{-6}$\\
		\hline
		15
		&13&$5.7\times 10^{-6}$          &13&$5.7\times 10^{-6}$\\
		\hline
		20
		&11&$1.3\times 10^{-5}$          &11&$1.3\times 10^{-5}$\\
		\hline
	\end{tabular}
	\label{tab4}%
\end{table}%

As demonstrated in Fig. \ref{O4}, for UAVs with different degrees of stability (i.e., UAVs with different variance values of orientation fluctuations), the optimal value of $N$ will be different. Moreover, as expected, in addition to the orientation fluctuations, the boresight direction of the antennas (i.e., $\theta'_{\textrm{ty}}$ and $\theta'_{\textrm{ry}}$) affects the optimal value of $N$. In particular, as an essential prerequisite to maintain the link alignment between transceivers, the instantaneous positions of UAVs and antenna boresight directions  should be accurately estimated  through information exchange between them.
Therefore, the angular offset, i.e., the difference between the actual value and the estimated value of boresight direction is another parameter for evaluating system performance.

To broaden our understanding of the impact of angular offset on the optimal value of $N$, in Figs. \ref{3D_uu} and \ref{3D_uu2}, the outage probability of a U2U link versus $N$ and $\sigma_{\textrm{ty}}$ is shown for two different angular offsets. Without loss of generality, in these simulations, $\sigma_{\textrm{ty}}$ and $\sigma_{\textrm{ry}}$ are assumed to be equal.
Note that, transceivers which employ  an antenna array with a high gain and narrow beam width are more vulnerable to the instantaneous orientation deviations of UAVs. Therefore, to increase the robustness of the system when the instantaneous orientation deviations are large, it is necessary to employ an antenna pattern with large beam width.
Meanwhile, the angular offset adversely affects the performance of the system as well as required number of $N$ to achieve target outage probability. On the other hand, when the transmit pattern experiences large AoD and AoA fluctuations, i.e., at large values of $\sigma_{\textrm{ty}}$ and $\sigma_{\textrm{ry}}$, increasing number of antenna elements leads to a narrower transmit beam and, in turn, an increase in the beam misalignment. Hence, a floor can be noticed for the outage probability.

Finally, to confirm the accuracy of our derived analytical expressions, in Tables \ref{tab2}-\ref{tab4}, the optimal number of $N$ and the corresponding outage probabilities obtained from both analytical and simulation results are shown for different values of orientation deviations. More precisely, Tables \ref{tab2} and \ref{tab3} show the results for the U2U link and the U2U2U link, respectively. Also, the results with taking  different values of angular offset into accounts are provided in Table \ref{tab4}.
For instance, from this table, by increasing the angular offset from 5 to 20 $\rm mrad$, the optimal $N$ decreases from 17 to 12 for $\textrm{SNR}=20$\,dB, and from 15 to 11 for $\textrm{SNR}=30$\,dB, respectively. As we observe from Tables \ref{tab2}-\ref{tab4}, the optimal value of antenna elements, $N_\textrm{opt}$, depends on both the angular offset and the variance of angular deviations. Since these two parameters may change during the aerial operation of a hovering UAV, the mounted ULA must be designed for the largest number of antenna (e.g., $N=18$ in our setup). Accordingly, under different conditions, only $N_\textrm{opt}$ of them are active to bring a reliable energy-efficient airborne system with long endurance. Meanwhile, the results of these tables confirm the accuracy of the proposed analytical expressions that makes it easy to study and design such UAV-based mmWave communication links.

\section{Conclusion }
{
In this paper, we have studied the problem of integrating mmWave frequencies on UAVs for providing wireless connectivity. Accordingly, we have considered three UAV-based mmWave communication links, namely U2U link, U2U2U link, and G2U2G link for which we have derived accurate and computationally efficient channel models.  Our simulation results have demonstrated that, unlike ground communication links, in the hovering UAV-based communication systems, the degree of stability of the mounted antennas on the transceivers has a considerable impact on the performance of the system and increasing antenna directivity gain does not necessarily improve the system performance. Hence, in the presence of hovering fluctuations, optimizing antenna radiation pattern plays a key role in such systems.
Our  analytical results have made it possible to find the optimal antenna directivity gain for designing a reliable UAV-based mmWave communication link  under different levels of stability of UAVs without resorting to time-consuming simulations.
}

\balance 

\end{document}